\documentclass[
amsmath,
floats,
floatfix,
12pt,
nofootinbib,
tightenlines,
showpacs]{revtex4-1}
\usepackage{amsmath}
\usepackage{amssymb}
\usepackage{graphicx}
\newcommand{\ljump}{\big[\!\big[}
\newcommand{\rjump}{\big]\!\big]}
\newcommand{\lcurlyjump}{\{\!\{}
\newcommand{\rcurlyjump}{\}\!\}}
\newcommand{\domain}{\mathsf{D}}
\begin{document}

\title{Discontinuous Galerkin method for the spherically reduced\\ 
       BSSN system with second-order operators}
      \author{
              Scott E.~Field${}^{1,}$\footnote{
              {\tt Scott\_Field@brown.edu},
              ${}^\dagger\,{}${\tt Jan\_Hesthaven@brown.edu},
              ${}^\ddagger\,{}${\tt srlau@math.unm.edu},
              ${}^\S\,{}${\tt mroue@cita.utoronto.ca}},
              Jan~S.~Hesthaven${}^{2,\dagger}$,
              Stephen R.~Lau${}^{3,\ddagger}$, and
              Abdul H.~Mroue${}^{4,\S}$
              }
              \affiliation{
              ${}^1$Department of Physics, Brown
               University, Providence, RI 02912, USA\\
              ${}^2$Division of Applied Mathematics, Brown
               University, Providence, RI 02912, USA\\
              ${}^3$Mathematics and Statistics, University
               of New Mexico, Albuquerque, NM 87131, USA\\
              ${}^4$Canadian Institute for Theoretical Astrophysics, 
                    University of Toronto, Toronto, 
                    Ontario M5S 3H8, Canada
             }
%%%%%%%%%%%%%%%%%%%%%%%%%%%%%%%%%%%%%%%%%%%%%%%%%%%%%%%%%%%%%%%
\begin{abstract}

We present a high-order accurate discontinuous 
Galerkin method for evolving the spherically-reduced 
Baumgarte-Shapiro-Shibata-Nakamura (BSSN) system expressed in 
terms of second-order spatial operators. Our multi-domain method achieves 
global spectral accuracy and long-time stability on short 
computational domains. We discuss in detail both our scheme
for the BSSN system and its implementation. After a 
theoretical and computational verification of the proposed 
scheme, we conclude with a brief discussion of issues likely 
to arise when one considers the full BSSN system.
\end{abstract}
\maketitle
%%%%%%%%%%%%%%%%%%%%%%%%%%%%%%%%%%%%%%%%%%%%%%%%%%%%%%%%%%%%%%%
%
%  INTRODUCTION
%
%%%%%%%%%%%%%%%%%%%%%%%%%%%%%%%%%%%%%%%%%%%%%%%%%%%%%%%%%%%%%%%
\section{Introduction}
\label{Sec:Introduction}

Breakthroughs in numerical relativity during this decade have
made it possible to simulate, via evolution of the full 3D 
Einstein equations, binary black hole dynamics through inspiral, 
merger and ringdown of the remnant single black hole~\cite{
Pretorius2005a,Pretorius2006,Campanelli2006a,Baker2006a,
Campanelli2006,Herrmann2007b,Diener2006,Scheel2006,Sperhake2006,
Bruegmann2006,Marronetti2007,Etienne2007,Szilagyi2007,
Boyle2007,Scheel2009}
(see e.~g.~recent reviews~\cite{Hannam:2009rd,Hinder:2010vn}).
Inspiraling binaries are among the most promising sources of
gravitational waves for the network of laser interferometric
detectors such as LIGO~\cite{LIGO} and VIRGO~\cite{VIRGO,VIRGO2}.
Through the construction of templates for matched filtering, 
waveforms extracted from numerical-relativity simulations 
are expected to facilitate the detection of genuine 
gravitational waveforms by interferometric detectors.

Early attempts to evolve the Einstein equations relied on 
the Arnowitt-Deser-Misner (ADM) decomposition~\cite{ADM,YorkSmarr}.
The resulting ADM system proved only weakly hyperbolic 
when expressed in first-order form, a fact partly accounting 
for difficulties associated with its numerical 
evolution~\cite{KST2001,CalabreseEtal2002}. Difficulties in 
evolving black-hole solutions to the Einstein equations 
also stem from singularities, gauge conditions within the 
computational domain, and unstable constraint violation.
For over ten years, the goal of accurate and stable 
numerical integration of the Einstein equations has 
continuously spurred the interest of numericists
and theorists alike, leading to a wealth of new 
formalisms 
\cite{frittelli94,
ShibataNakamura95,
choquet_york95,
abrahams_etal95,
bona_masso95b,
mvp96,
frittelli_reula96,
Friedrich96,
estabrook_etal97,
Iriondo1997,
anderson_etal98,
Bonilla1998,
Baumgarte99,
Yoneda1999,
Alcubierre1999,
Frittelli1999,
anderson_york99,
Hern1999,
Friedrich_Rendall2000,
Yoneda2000,
Shinkai2000,
Yoneda2001,
Kidder2001,
GundlachelAL2005,
Pretorius2005c,
BrownGBSSN,
Lindblom2006,
Owen2007}
(this list is not exhaustive).

To evolve binary black holes, numerical relativists currently 
use one of the following versions of the Einstein equations: 
the generalized harmonic (GH) system 
\cite{Fock,GundlachelAL2005,Pretorius2005c,Lindblom2006}
or the Baumgarte-Shapiro-Shibata-Nakamura (BSSN) system
\cite{ShibataNakamura95,Baumgarte99,BrownGBSSN}.
Using a finite-difference approach with adaptive mesh
refinement, Pretorius~\cite{Pretorius2005a,Pretorius2005c,Pretorius2006} 
used a constraint-suppressing second-order form of the GH 
system (suggested by Gundlach et al.~\cite{GundlachelAL2005}) 
to evolve a binary through inspiral, merger and ringdown. 
Lindblom et al.~\cite{Lindblom2006} recast the second-order 
GH system into a first-order symmetric-hyperbolic evolution 
system with constraint suppression comparable to that of 
the second-order system. This first-order GH system has 
been used to successfully simulate binary black holes 
evolution with nodal spectral (pseudospectral) 
methods~\cite{Boyle2007,Scheel2009,SpEC}. More recently, 
Ref.~\cite{NTaylor} has introduced a new penalty method for 
nodal spectral evolutions of spatially second-order wave 
equations. This work provides a foundation for solution of 
the second-order GH system via spectral methods, and has 
been used to evolve the Kerr solution \cite{NTaylor2}
and the inspiral of binaries.
Typically written in a spatially second-order form, the BSSN 
system~\cite{Baumgarte99} has seen widespread use by numerical 
relativity groups that employ finite-difference techniques to
evolve binaries. Ref.~\cite{Tichy2006} presented a nodal spectral 
code to evolve the BSSN system in second-order form. The system 
proved unstable when tested on a single black hole. In more 
recent work \cite{Tichy2009}, longer evolutions were obtained 
through the adoption of better gauge conditions, filtering methods,
and more distant outer boundaries. The BSSN system has also been 
evolved in a first-order strongly-hyperbolic formulation for a 
single black hole with nodal spectral methods~\cite{Mroue}.
Such evolutions of a single black hole exhibited instabilities
similar to those reported in Ref.~\cite{Tichy2009}.

Corresponding to the two versions of the Einstein 
equations discussed in the last paragraph are two distinct 
techniques for the treatment of singularities in numerical 
relativity. Evolutions based on the GH system have used 
black-hole excision, whereby the interior of an apparent 
horizon is removed (excised) from the computational domain. 
This technique relies on horizon-tracking and gauge 
conditions which ensure that inner boundaries of the 
computational domain are pure out-flow, whence no inner 
boundary conditions are needed. Evolutions based on the 
BSSN system have relied on the moving-punctures 
technique~\cite{Campanelli2006a,Baker2006a}, also 
coined ``natural excision." Technically much easier to 
implement than excision, this technique features mild 
central singularities which evolve freely in the 
computational domain. Initially these puncture points 
may represent either asymptotically flat regions or 
``trumpets." Hannam et al.~first discussed 
cylindrical asymptotics in moving puncture evolutions 
\cite{HannamPuntures1,HannamPuntures2}, see also
\cite{BrownSphSymBSSN,
BaumgarteTrumpet,
HannamTrumpet,
BrownPuncture1,
BrownPuncture2}.

Relative to the alternative systems previously discussed, 
the BSSN system in second order form affords an easier 
treatment of singularities and features a relatively small number 
of geometric variables directly related to the foliation of spacetime 
into spacelike hypersurfaces. However, to date, spectral 
methods for black-hole binaries have been successfully implemented 
only for the first-order GH system. The binary black hole problem 
is essentially a smooth one (singularities reside on sets of measure 
zero censored by horizons), and spectral methods exhibit 
well-established advantages over finite-difference methods for 
long-time simulation of such problems \cite{HEST_GOT}. Therefore, 
the development and analysis of a stable spectral implementation 
of the full BSSN system is a worthwhile goal in numerical 
relativity, and the motivation behind the pioneering investigations 
reported in Refs.~\cite{Tichy2006,Tichy2009,Mroue}.

In Refs.~\cite{BrownGBSSN,BrownSphSymBSSN}, Brown introduced 
a spherically reduced version of the BSSN system as a test bed
for tractable examination of theoretical and computational 
issues involved in solving this system. 
Indeed, appealing to the simplicity of this system, 
he offered geometrical and physical insights into the nature 
of the moving-puncture technique and its finite-difference 
implementation \cite{BrownSphSymBSSN,BrownPuncture1,BrownPuncture2} 
(see also \cite{BaumgarteTrumpet,HannamTrumpet}). 
Here, we exploit this system to a similar end, using it as
a simplified setting in which to develop spectral methods for
the stable integration of the BSSN system. Precisely, we develop 
and test a nodal discontinuous Galerkin method (dG) \cite{HEST} 
for integration of the spherically reduced BSSN system. 
While Brown's chief focus lay with moving punctures, for 
further simplicity we adopt the excision technique. Clearly,
the problem we consider is not as daunting as the one confronted 
by both Tichy and Mroue \cite{Tichy2006,Tichy2009,Mroue}. 
Nevertheless, our method is robustly stable, and therefore might 
serve as a stepping stone toward a stable dG-based formulation
for the full BSSN system. The conclusion offers further 
comments toward this end.

Nodal dG schemes are both well-suited and well-developed 
for hyperbolic problems \cite{HEST}. Although mostly used 
for hyperbolic problems expressed as first-order 
systems, dG methods have also been applied to systems 
involving second-order spatial operators, typically via
dG {\em interior penalty} (IP) methods
\cite{
Shahbazi_IP,
ShuAltdG,
GroteWave1,
GroteWave2,
HesthavenWarburtonMaxwell,
SchneebeliThesis}. 
(Refs.~\cite{Friedrich,Gundlach_2004,Gundlach_2006}
discuss the concept of hyperbolicity
\cite{KreissLorenz} in the context of such systems.)
Penalty methods of a different type were exploited 
in Ref.~\cite{NTaylor} for the wave equation written in 
second order form.
{\em Local discontinuous Galerkin} (LDG) schemes,
developed initially by Shu and coworkers 
\cite{CockburnShu,ShuSchro,ShuKvd}, constitute an alternate 
approach for integration of spatially second-order
systems. LDG schemes feature essentially the same 
auxiliary variables as those appearing in traditional 
first-order reductions, however in LDG schemes such 
variables are not evolved and arise only as local 
variables. The basic difference between dG--IP 
and LDG methods is the manner in which subdomains are coupled. 
The method we described for the spherically reduced BSSN system is 
essentially an LDG scheme.

This paper is organized as follows. Section \ref{sec:GBSSN}
collects the relevant equations from Brown's presentation, 
and develops some further notation useful for expressing
the spherically reduced BSSN system in various abstract forms.
Section \ref{sec:DG} presents our nodal dG scheme in detail, 
and Section \ref{sec:NumSims} documents the results of 
several numerical simulations testing our scheme. Our 
conclusion discusses possible generalization of our method 
to the full BSSN system. Several appendices collect further 
technical details. In particular, Appendix \ref{WPToy} 
considers a simple system which models the spherically 
reduced BSSN system, giving an analytical proof that the 
model system is $L_2$ stable in the semi-discrete sense.

%%%%%%%%%%%%%%%%%%%%%%%%%%%%%%%%%%%%%%%%%%%%%%%%%%%%%%%%%%%%%%%
%
%  SECTION 2: GBSSN Equations
%
%%%%%%%%%%%%%%%%%%%%%%%%%%%%%%%%%%%%%%%%%%%%%%%%%%%%%%%%%%%%%%%
\section{Spherically symmetric (generalized) 
BSSN equations}\label{sec:GBSSN}

As shown by Brown \cite{BrownGBSSN}, the BSSN system can 
be generalized to allow for a conformal metric without unit 
determinant, and this paper focuses on the spherical 
reduction of this system, also considered by Brown in 
\cite{BrownSphSymBSSN}. In fact, this spherical 
reduction relies on freedom present in the generalized 
BSSN system, since spherical-polar coordinates should not be 
associated with a unit-determinant conformal metric. 
Although we work with the spherically 
reduced {\em generalized} BSSN system (subject to Brown's
Lagrangian condition, to be precise), we will nevertheless 
describe it as the spherically reduced BSSN system.

\subsection{Basic variables and spherically reduced system.}
The conformal-traceless decomposition of the geometry associated
with a spacelike 3-surface is
\begin{equation}\label{eqn:conMet}
\bar{g}_{ab} = \chi^{-1} g_{ab},\qquad
K_{ab} = \chi^{-1}\big(A_{ab} + \frac{1}{3}g_{ab}K\big),
\end{equation}
where $\bar{g}_{ab}$ is the physical 3-metric and $K_{ab}$
is the physical extrinsic curvature tensor. The BSSN variables
are the conformal metric $g_{ab}$, the conformal factor $\chi$,
the trace-free extrinsic curvature $A_{ab}$, the trace 
$K =\bar{g}{}^{ab}K_{ab}$, and the conformal connection 
$\Gamma^a \equiv - g^{-1/2}\partial_b (g^{1/2} g^{ab})$,
where $g$ is the determinant of the metric. 
The BSSN system also includes the lapse 
$\alpha$, shift vector $\beta^a$, and an auxiliary vector
field $B^a$ used to define the ``$\Gamma$-driver" for the
shift.

Following Brown, we adopt a spherically symmetric line 
element,
\begin{equation}\label{eqn:metric}
ds^2 = - \alpha^2 dt^2 
       + \chi^{-1} g_{rr}(dr+\beta^r dt)^2
       + \chi^{-1} g_{\theta\theta}
         (d\theta^2 + \sin^2\theta d\phi^2),
\end{equation}
along with the spherically symmetric {\em Ansatz}: 
\begin{subequations}\label{eqn:Ansatz}
\begin{align}
\Gamma^a = \left(\begin{array}{c}\Gamma^r\\ 
            -\cos\theta /
             (g_{\theta\theta}\sin\theta)\\ 
             0
             \end{array}
             \right),
\quad
A_{ab} = A_{rr}
             \left(\begin{array}{ccc}
             1 & 0 & 0 \\ 
             0 & -g_{\theta\theta}/(2g_{rr}) & 0 \\
             0 & 0 & -g_{\theta\theta}\sin^2\theta/(2g_{rr}) 
             \end{array}
       	     \right).
\end{align}
\end{subequations}
Subject to the assumption of spherical symmetry, the basic variables
are $\chi$, $g_{rr}$, $g_{\theta\theta}$, $A_{rr}$, $K$, $\Gamma^r$,
$\alpha$, $\beta^r$, $B^r$. All are functions of $t$ and 
$r$, and satisfy the following spherically symmetric 
(generalized, Lagrangian-form) BSSN system:\footnote{For 
this system the determinant 
$g = g_{rr}(g_{\theta\theta})^2\sin^4\theta$ is not unity}.
\begin{subequations}\label{eqn:GBSSNL_spherical}
\begin{align}
\partial_t \alpha           & = \beta^r\alpha' 
                               -2\alpha K 
                               -(\partial_t\alpha)_0
\\
\partial_t \beta^r          & = \beta^r {\beta^r}'
                               +\frac{3}{4}B^r 
                               -(\partial_t\beta^r)_0 
\\
\partial_t B^r              & = \beta^r {B^r}'
                               +\lambda
                                (\partial_t \Gamma^r 
                               -\beta^r {\Gamma^r}') 
                               -\eta B^r 
                               -(\partial_t B^r)_0 
\\
\partial_t \chi             & = \beta^r \chi' 
                               +\frac{2}{3}K\alpha\chi 
                               -\frac{\beta^r g_{rr}'\chi}{3g_{rr}}
                               -\frac{2\beta^r 
                                      g_{\theta \theta}'
                                      \chi}{3g_{\theta\theta}} 
                               -\frac{2}{3}{\beta^r}'\chi
\\
\partial_t g_{rr}           & = \frac{2}{3}\beta^r g_{rr}' 
                               +\frac{4}{3}g_{rr}{\beta^r}' 
                               -2A_{rr}\alpha 
                               -\frac{2g_{rr}\beta^r 
                                      g_{\theta\theta}'}{
                                      3g_{\theta\theta}}
\\
\partial_t g_{\theta\theta} & = \frac{1}{3}\beta^r g_{\theta\theta}' 
                               +\frac{A_{rr}g_{\theta\theta}
                                      \alpha}{g_{rr}}
                               -\frac{g_{\theta\theta}
                                      \beta^r g_{rr}'}{3g_{rr}} 
                               -\frac{2}{3}g_{\theta\theta}
                                {\beta^r}'
\\
\partial_t A_{rr}           & = \beta^r A_{rr}'
                               +\frac{4}{3}A_{rr}{\beta^r}' 
                               -\frac{\beta^r g_{rr}' 
                                      A_{rr}}{3g_{rr}} 
                               -\frac{2\beta^r g_{\theta\theta}' 
                                      A_{rr}}{3g_{\theta\theta}} 
                               +\frac{2\alpha \chi 
                                      (g_{rr}')^2}{3g_{rr}^2} 
                               -\frac{\alpha \chi 
                                      (g_{\theta \theta}')^2}{
                                      3g_{\theta\theta}^2} 
                               -\frac{\alpha (\chi ')^2}{
                                      6\chi} 
\nonumber \\
                            &  +\frac{2}{3}g_{rr}\alpha\chi{\Gamma^r}' 
                               -\frac{\alpha\chi g_{rr}'
                                      g_{\theta\theta}'}{
                                      2g_{rr}g_{\theta \theta}} 
                               +\frac{\chi g_{rr}' 
                                      \alpha'}{3g_{rr}}
                               +\frac{\chi g_{\theta\theta}'
                                      \alpha'}{3g_{\theta\theta}}
                               -\frac{\alpha g_{rr}' 
                                      \chi'}{6g_{rr}}
                               -\frac{\alpha g_{\theta\theta}' 
                                      \chi'}{6g_{\theta\theta}} 
                               -\frac{2}{3}\alpha' \chi'
                               +\frac{\alpha\chi''}{3}
\nonumber \\
                            &  -\frac{2}{3}\chi \alpha''
                               -\frac{\alpha\chi g_{rr}''}{3g_{rr}} 
                               +\frac{\alpha\chi 
                                     g_{\theta\theta}''}{3g_{\theta\theta}}  
                               -\frac{2\alpha A_{rr}^2}{g_{rr}} 
                               +K\alpha A_{rr} 
                               -\frac{2g_{rr}\alpha 
                                     \chi}{3g_{\theta\theta}} 
\\
\partial_t K                & = \beta^r K' 
                               +\frac{\chi g_{rr}'\alpha'}{2g_{rr}^2} 
                               -\frac{\chi g_{\theta\theta}'
                                      \alpha'}{g_{rr}g_{\theta\theta}} 
                               +\frac{\alpha'\chi'}{2g_{rr}} 
                               -\frac{\chi\alpha''}{g_{rr}} 
                               +\frac{3\alpha A_{rr}^2}{2g_{rr}^2}
                               +\frac{1}{3}\alpha K^2
\\
\partial_t \Gamma^r         & = \beta^r {\Gamma^r}'
                               +\frac{A_{rr}\alpha 
                                      g_{\theta\theta}'}{g_{rr}^2 
                                      g_{\theta \theta}} 
                               +\frac{2 {\beta^r}'g_{\theta \theta}'}{
                                      3g_{rr}g_{\theta\theta}} 
                               +\frac{A_{rr}\alpha g_{rr}'}{g_{rr}^3} 
                               -\frac{4\alpha K'}{3g_{rr}}
                               -\frac{2A_{rr}\alpha '}{g_{rr}^2}
                               -\frac{3A_{rr}\alpha \chi '}{g_{rr}^2 \chi} 
\nonumber \\
                            & + \frac{4{\beta^r}''}{3g_{rr}}
                              -\frac{\beta^r (g_{\theta\theta}')^2}{
                                     g_{rr}(g_{\theta\theta})^2} 
                              +\frac{\beta^r g_{rr}''}{6(g_{rr})^2}
                              +\frac{\beta^r g_{\theta \theta}''}{
                                     3g_{\theta\theta}g_{rr}},
\end{align}
\end{subequations}
where the prime stands for partial $r$-differentiation.
Eqs.~(\ref{eqn:GBSSNL_spherical}d-i) are Brown's Eqs.~(9a-f) 
listed in \cite{BrownSphSymBSSN}, subject to his Lagrangian 
condition (corresponding to $v = 1$ in Brown's equations). 
The first three equations 
(\ref{eqn:GBSSNL_spherical}a-c) comprise the gauge sector, and
these are essentially spherically symmetric versions of the
standard ``1+log" and ``$\Gamma$-driver" conditions listed in 
Eqs.~(1) and (2) of \cite{BrownSphSymBSSN}. However, we have 
introduced the following minor modifications. First,
$(\partial_t\alpha)_0$ designates a constant term which ensures that 
the right-hand side of the $\alpha$ evolution equation 
(\ref{eqn:GBSSNL_spherical}a) vanishes at the initial time.
This source term as well as the analogous terms appearing in
the evolution equations (\ref{eqn:GBSSNL_spherical}b,c) for 
$\beta^r$ and $B^r$ are needed to enable a static evolution of the 
Schwarzschild solution in Kerr-Schild coordinates. 
Second, the parameter $\lambda$ (perhaps with functional 
dependence) modifies the hyperbolicity of the first-order 
system \cite{Alcubierre2003}. The damping parameter $\eta$ typically 
appears in standard versions of these gauge evolution equations. 
(See Sections \ref{subsec:hyperbolicity-char-fields} and 
~\ref{subsec:Results_KerrSchild} for further discussions.)
For this BSSN system, we have three constraints: 
the Hamiltonian constraint $\mathcal{H}$, the 
momentum constraint $\mathcal{M}_r$, and the 
constraint $\mathcal{G}^r$ resulting from the 
definition of the conformal connection $\Gamma^r$. 
In spherical symmetry, these constraints are 
written as follows:
\begin{subequations}
\begin{align}
\mathcal{H} & = -\frac{3A_{rr}^2}{2g_{rr}^2} 
                +\frac{2K^2}{3}
                -\frac{5(\chi')^2}{2\chi g_{rr}} 
                +\frac{2\chi''}{g_{rr}}
                +\frac{2\chi}{g_{\theta\theta}} 
                -\frac{2\chi g_{\theta\theta}''}{g_{rr}g_{\theta\theta}}
                + \frac{2\chi'g_{\theta\theta}'}{g_{rr}g_{\theta\theta}}
                + \frac{\chi g_{rr}'g_{\theta\theta}'}{g_{rr}^2g_{\theta\theta}}
                - \frac{\chi'g_{rr}'}{g_{rr}^2}
                + \frac{\chi (g_{\theta\theta}')^2}{2g_{rr}g_{\theta\theta}^2}
\\
\mathcal{M}_r & = \frac{A_{rr}'}{g_{rr}} 
                 -\frac{2 K'}{3}
                 -\frac{3 A_{rr}\chi'}{2\chi g_{rr}}
                 +\frac{3 A_{rr}g_{\theta\theta}'}{2g_{rr}g_{\theta\theta}}
                 -\frac{A_{rr}g_{rr}'}{g_{rr}^2}
\\
\mathcal{G}^r & = -\frac{g_{rr}'}{2g_{rr}^2} 
                  +\frac{g_{\theta\theta}'}{g_{rr}g_{\theta\theta}}
                  +\Gamma^r.
\end{align}
\end{subequations}
These expressions are the ones listed by Brown in \cite{BrownSphSymBSSN}.
Eqs.~(\ref{eqn:GBSSNL_spherical}e,f) also ensure that the
determinant factor $g/\sin^4\theta = g_{rr} (g_{\theta\theta})^2$ 
remains fixed throughout an evolution.

\subsection{Abstract expressions of the system}
We define the following vectors built with 
system variables:
\begin{equation}\label{eqn:vectors_uvQ}
u = \left(
\begin{array}{c}
\chi\\
g_{rr}\\
g_{\theta\theta}\\
\alpha\\
\beta^r
\end{array}
\right),\qquad
v = \left(
\begin{array}{c}
B^r\\
A_{rr}\\
K\\
\Gamma^r
\end{array}
\right),\qquad
Q = \left(
\begin{array}{c}
\chi '\\
g_{rr} '\\
g_{\theta\theta} '\\
\alpha '\\
{\beta^r}'
\end{array}
\right).
\end{equation}
Introduction of $Q$ might seem unnecessary at this 
stage, but proves useful in the construction of our
discontinuous Galerkin scheme. In terms of the vectors
$u$, $v$, and $Q$ we further define
\begin{equation}\label{eqn:systemvectors}
W_{u:v} = \left(
\begin{array}{c}
u\\
v
\end{array}
\right),\qquad
W_{v:Q} = \left(
\begin{array}{c}
v\\
Q
\end{array}
\right),\qquad
W = W_{u:Q}
= \left(
\begin{array}{c}
u\\
v\\
Q
\end{array}
\right).
\end{equation}
Here we have introduced ``colon notation" \cite{GolubVanLoan}
to represent (sub)vectors and (sub)matrices, although
we employ the notation over block rather than individual
elements. In the first-order version of the system 
\eqref{eqn:GBSSNL_spherical}
the components of $Q$ are promoted to independent fields, in which 
case the corresponding principal part features
\begin{subequations}\label{eqn:GBSSNPrinciplePart}
\begin{align}
\partial_t B^r              & = \beta^r {B^r}'
                               -\frac{4\lambda\alpha}{3g_{rr}}K'
                               + \frac{4\lambda}{3g_{rr}}Q_{\beta^r}'
                              +\frac{\lambda\beta^r}{6(g_{rr})^2}
                               Q_{g_{rr}}'
                              +\frac{\lambda\beta^r}{3g_{\theta\theta}g_{rr}}
                               Q_{g_{\theta\theta}}'
\\
\partial_t A_{rr}           & = \beta^r A_{rr}'
                               +\frac{2}{3}g_{rr}\alpha\chi{\Gamma^r}' 
                               +\frac{1}{3}\alpha Q_\chi'
                               -\frac{2}{3}\chi Q_\alpha'
                               -\frac{\alpha\chi}{3g_{rr}}
                               Q_{g_{rr}}' 
                               +\frac{\alpha\chi}{3g_{\theta\theta}}  
                               Q_{g_{\theta\theta}}'
\\
\partial_t K                & = \beta^r K' 
                               -\frac{\chi}{g_{rr}}Q_\alpha' 
\\
\partial_t \Gamma^r         & = \beta^r {\Gamma^r}'
                               -\frac{4\alpha K'}{3g_{rr}}
                               +\frac{4}{3g_{rr}}Q_{\beta^r}'
                               +\frac{\beta^r}{6(g_{rr})^2}
                               Q_{g_{rr}}'
                               +\frac{\beta^r}{3g_{\theta\theta}g_{rr}}
                               Q_{g_{\theta\theta}}'
\\
\partial_t Q_\chi
                            & = \beta^r Q_\chi'
                               +\frac{2}{3}\alpha\chi K'
                               -\frac{\beta^r\chi}{3g_{rr}}
                                Q_{g_{rr}}'
                               -\frac{2\beta^r\chi}{3g_{\theta\theta}}
                                Q_{g_{\theta\theta}}'
                               -\frac{2}{3}\chi Q_{\beta^r}'
\\
\partial_t Q_{g_{rr}}
                            & = \frac{2}{3}\beta^r Q_{g_{rr}}'
                               +\frac{4}{3}g_{rr}Q_{\beta^r}'
                               -2\alpha A_{rr}'
                               -\frac{2g_{rr}\beta^r}{3g_{\theta\theta}}
                                Q_{g_{\theta\theta}}'
\\
\partial_t Q_{g_{\theta\theta}}
                           & = \frac{1}{3}\beta^r Q_{g_{\theta\theta}}'
                               +\frac{g_{\theta\theta}
                                      \alpha}{g_{rr}} A_{rr}'
                               -\frac{g_{\theta\theta}\beta^r}{3g_{rr}}
                                Q_{g_{rr}}'
                               -\frac{2}{3}g_{\theta\theta}
                                Q_{\beta^r}'
\\
\partial_t Q_\alpha
                            & = \beta^r Q_\alpha'
                               -2\alpha K'
\\
\partial_t Q_{\beta^r}
                            & = \beta^r Q_{\beta^r}'
                               +\frac{3}{4}{B^r}',
\end{align}
\end{subequations}
where all lower-order terms on the right-hand side have been dropped.
This sector of principal parts of the first-order system has the form
\begin{equation}\label{eqn:partofpp}
\partial_t W_{v:Q} + \tilde{A}(u) W_{v:Q}' = 0,
\end{equation}
where (minus) the explicit form of the 9-by-9 matrix $\tilde{A}(u)$
is given below in \eqref{eqn:matrix_pp}. The first-order
version of \eqref{eqn:GBSSNL_spherical} takes the 
nonconservative form
\begin{equation}\label{eqn:firstordersys_abs}
\partial_t W + \mathcal{A}(u) W' = \mathcal{S}(W),
\qquad
\mathcal{A}(u) = \left(
\begin{array}{c|c}
0_{5 \times 5} & 0_{5\times 9}\\
\hline
0_{9\times 5}  & \overset{}{\tilde{A}(u)}
\end{array}
\right),
\end{equation}
where $\mathcal{S}(W)$ is a vector of lower order terms
built with all components of $W$. Partition of
$\tilde{A}(u) = \mathcal{A}(u)_{v:Q,v:Q}$ into blocks 
corresponding to the $v$ and $Q$ sectors yields
\begin{equation}
\tilde{A}(u) = \left(
\begin{array}{c|c}
\tilde{A}(u)_{vv} & \tilde{A}(u)_{vQ}\\
\hline
\overset{}{\tilde{A}(u)_{Qv}} & \overset{}{\tilde{A}(u)_{QQ}}
\end{array}
\right).
\end{equation}
Using these blocks, we then define the 9-by-9 matrix
\begin{equation}\label{eqn:Aofymatrix}
A(u) = \mathcal{A}(u)_{u:v,v:Q} =
\left(
\begin{array}{c|c}
0_{5\times 4} 
&
0_{5\times 5}\\
\hline
\overset{}{\tilde{A}(u)_{vv}}
&
\overset{}{\tilde{A}(u)_{vQ}}
\end{array}
\right),
\end{equation}
and express \eqref{eqn:GBSSNL_spherical} as
\begin{subequations}\label{eqn:NonconservativeForm}
\begin{align}
\partial_t W_{u:v} + A(u) W_{v:Q}' & = S(W) \\
           Q & = u',
\end{align}
\end{subequations}
where $S(W) = \mathcal{S}(W)_{u:v}$. 

\subsection{Hyperbolicity and characteristic fields}
\label{subsec:hyperbolicity-char-fields}
Although our numerical scheme deals directly with the
second-order spatial operators appearing in 
\eqref{eqn:GBSSNL_spherical}, we first consider the
hyperbolicity of the corresponding first-order system
\eqref{eqn:firstordersys_abs}. The characteristic fields 
and their speeds are found by instantaneously ``freezing" 
the fields $u$ in $\mathcal{A}(u)$ to some value $u_0$,
corresponding to a linearization around a uniform state. 
Below we continue to write $u$ for simplicity with the 
understanding that $u$ is really the background solution 
$u_0$. Of primary interest is the range of $u_0$ for
which the system is strongly hyperbolic 
\cite{Friedrich,
Gundlach_2004,
Gundlach_2006,
KreissLorenz}.

\begin{table}
\begin{tabular}{|l||l|}
\hline
$\;$ field $\;$    & $\;$ speed \\
\hline
\hline
$X_1$              & $\mu_1 = 0$                                             \\ \hline 
$X_{2,3}$          & $\mu_{2,3} = -\beta^r$                                  \\ \hline
$X^\pm_4$          & $\mu^\pm_4 = -\beta^r \pm \sqrt{2\alpha\chi/g_{rr}}$    \\ \hline
$X^\pm_5$          & $\mu^\pm_5 = -\beta^r \pm \alpha \sqrt{\chi/g_{rr}}$    \\ \hline
$X^\pm_6$          & $\mu^\pm_6 = -\beta^r \pm \sqrt{\lambda/g_{rr}}$        \\ \hline
\end{tabular}
\caption{{\sc Characteristic speeds.} These speeds are the 
eigenvalues listed in \eqref{eqn:mu_evalues}.
\label{tab:speeds}}
\end{table}

Appendix \ref{app:hyperbolic} shows that the characteristic fields 
corresponding to \eqref{eqn:GBSSNL_spherical} are as follows: (i)
all components of $u$ (each with speed 0), and (ii) the fields
\begin{subequations}\label{eqn:characteristic_fields}
\begin{align}
X_1 & = g_{\theta \theta}Q_{{g}_{rr}}+2g_{rr}Q_{{g}_{\theta \theta}}
\\
X_2 & = 
g_{rr}\Gamma^r 
+\frac{2}{\chi}Q_{{\chi}}
-\frac{1}{2g_{rr}}Q_{{g}_{rr}}
-\frac{1}{g_{\theta \theta}}Q_{{g}_{\theta \theta}}
\\
X_3 & = \frac{g_{rr}}{\lambda}B^r 
+\frac{2}{\chi}Q_\chi
-\frac{1}{2g_{rr}}Q_{{g}_{rr}}
-\frac{1}{g_{\theta \theta}}Q_{g_{\theta \theta}}
\\
X_4^\pm & = \pm \sqrt{\frac{2\alpha g_{rr}}{\chi}}K +  Q_{\alpha}
\\
X_5^\pm & = 
\mp \frac{3}{\sqrt{g_{rr}\chi}}A_{rr}
\pm 2\sqrt{\frac{g_{rr}}{\chi}}K
+2g_{rr}\Gamma^r
+\frac{1}{\chi}Q_{\chi}
-\frac{1}{g_{rr}}Q_{g_{rr}}
+\frac{1}{g_{\theta \theta}}Q_{{g}_{\theta \theta}}\\
X_6^\pm & = 
-\frac{3}{4}\frac{g_{rr}}{\lambda}B^r
\pm  \frac{\alpha\sqrt{\lambda g_{rr}}}{(2\alpha \chi-\lambda)} K
-\frac{\beta^r}{8(\beta^r g_{rr} \mp \sqrt{\lambda g_{rr}})}Q_{{g}_{rr}}
\nonumber \\
& 
-\frac{\beta^r g_{rr}}{4g_{\theta \theta}(\beta^r g_{rr} 
 \mp \sqrt{\lambda g_{rr}})}Q_{{g}_{\theta \theta}}
+\frac{\alpha \chi}{(2\alpha \chi -\lambda)}Q_{{\alpha}}
\pm \sqrt{\frac{g_{rr}}{\lambda}}Q_{{\beta^r}},
\end{align}
\end{subequations}
with the speeds listed in Table \ref{tab:speeds}. 
To ensure strong hyperbolicity
we must necessarily require
\begin{equation}\label{eqn:hyperbolicity_cond}
\lambda > 0,                         \qquad 
(\beta^r)^2 g_{rr} - \lambda \neq 0, \qquad 
2\alpha\chi - \lambda \neq 0,
\end{equation}
as shown in in Appendix
\ref{app:hyperbolic} where further conditions are
also given.
When $\lambda = 1$ the hyperbolicity condition
of Ref.~\cite{BrownSphSymBSSN} is recovered. 
In fact, the system could be recast as symmetric hyperbolic.
Indeed, as it involves one spatial dimension, the relevant 
symmetrizer can be constructed via polar decomposition
of the diagonalizing similiarity transformation. However, we 
will not exploit this possibility.

This system admits an inner excision boundary 
provided
\begin{align}\label{eqn:excision_cond}
\beta^r  \geq \mathrm{max}
\left(
\sqrt{\frac{2\alpha \chi}{g_{rr}}}, 
\hspace{5pt} 
\sqrt{\frac{\alpha^2 \chi}{g_{rr}}}, 
\hspace{5pt} 
\sqrt{\frac{\lambda}{g_{rr}}}
\right)
\end{align}
holds at the inner boundary. This condition ensures 
each characteristic field has a nonpositive speed at
the inner boundary, and therefore the inner boundary
is an excision boundary at which no boundary conditions
are needed. The extra flexibility afforded by the parameter 
$\lambda$ could be used to maintain rigorous hyperbolicity by moving 
the points at which the conditions in 
\eqref{eqn:hyperbolicity_cond} are violated outside of 
the computational domain. Furthermore, for $\lambda = 1$
Eq.~\eqref{eqn:excision_cond} conceivably fails or is only 
satisfied close to $r=0$ where field gradients are 
prohibitively large. The troublesome $X_6^+$ gauge mode has a 
positive speed $-\beta^r + \sqrt{\lambda/g_{rr}}$. 
Indeed, for the conformally flat Kerr-Schild 
system considered in section \ref{subsec:Results_KerrSchild} 
an inner excision boundary is only possible provided 
$\lambda$ is small enough. 

The transformation \eqref{eqn:characteristic_fields} 
can be inverted in order to express the fundamental 
fields in terms of the characteristic fields:
\begin{subequations}\label{eqn:fundvarsFROMcharvars}
\begin{align}
B^r                  & = 
-\frac{1}{6} \frac{\lambda }{ g_{rr} g_{\theta\theta} }
\left[\frac{(\beta^r)^2}{(\beta^r)^2 g_{rr} - \lambda}\right]
X_1 
+ \frac{2}{3}
\frac{\lambda\alpha\chi}{g_{rr}(2\alpha\chi-\lambda)}
(X^+_4 + X^-_4)
-\frac{2}{3} \frac{\lambda}{g_{rr}}
(X^+_6 + X^-_6)
\\
A_{rr}               & = 
\frac{1}{3}\sqrt{\frac{g_{rr}\chi}{2\alpha}}
(X_4^+ - X_4^-)
- \frac{\sqrt{g_{rr}\chi}}{6}
(X_5^+ - X_5^-)
\\
K                    & = 
\sqrt{\frac{\chi}{8\alpha g_{rr}}}
(X_4^+ - X_4^-)
\\
\Gamma^r             & =
-\frac{1}{6} \frac{1}{ g_{rr} g_{\theta\theta} }
\left[\frac{(\beta^r)^2}{(\beta^r)^2 g_{rr} - \lambda}\right]
X_1
+ \frac{1}{g_{rr}}
(X_2 - X_3)
+ \frac{2}{3}
\frac{\alpha\chi}{g_{rr}(2\alpha\chi-\lambda)}
(X^+_4 + X^-_4)
\nonumber \\
& -\frac{2}{3} \frac{1}{g_{rr}}
(X^+_6 + X^-_6)
\\
Q_\chi               & =
\frac{1}{12} \frac{\chi}{g_{rr}g_{\theta\theta}}
\left[\frac{4(\beta^r)^2g_{rr} 
- 3\lambda}{(\beta^r)^2 g_{rr} - \lambda}\right]
X_1 
+ \frac{\chi}{2} 
X_3
- \frac{1}{3}\frac{\alpha\chi^2}{(2\alpha\chi-\lambda)}
(X^+_4 + X^-_4)
\nonumber
\\
& +\frac{\chi}{3}
(X^+_6 + X^-_6)
\\
Q_{g_{rr}}           & = 
\frac{2 (\beta^r)^2 g_{rr} - 3\lambda}{
6 g_{\theta\theta} ((\beta^r)^2 g_{rr} - \lambda)}
X_1
+\frac{4}{3}g_{rr}
X_2 
- g_{rr} 
X_3
+\frac{2}{3} \frac{\alpha\chi g_{rr}}{(2\alpha\chi -\lambda)}
(X^+_4 + X^-_4)
\nonumber \\
& -\frac{1}{3}g_{rr}
(X^+_5 + X^-_5)
  -\frac{2}{3}g_{rr}
(X^+_6 + X^-_6)
\\
Q_{g_{\theta\theta}} & = 
\left[\frac{1}{4g_{rr}} 
+ \frac{(\beta^r)^2}{12((\beta^r)^2 g_{rr}-\lambda)}\right]
X_1
-\frac{2}{3}g_{\theta\theta}
X_2
+ \frac{1}{2}g_{\theta\theta}
X_3
- \frac{1}{3}\frac{\alpha\chi 
g_{\theta\theta}}{(2\alpha\chi-\lambda)}
(X^+_4 + X^-_4) \nonumber 
\\
& + \frac{1}{6}g_{\theta\theta} 
(X^+_5 + X^-_5)
  + \frac{1}{3}g_{\theta\theta} 
(X^+_6 + X^-_6)
\\
Q_\alpha             & = \frac{1}{2}(X_4^+ + X_4^-)\\
Q_{\beta^r}          & = 
\frac{\beta^r \lambda}{8 g_{rr} g_{\theta\theta}
((\beta^r)^2 g_{rr}-\lambda)} X_1
-\frac{\lambda}{(2\alpha\chi-\lambda)}
\sqrt{\frac{\alpha\chi}{8g_{rr}}}
(X^+_4 - X^-_4)
+\frac{1}{2}\sqrt{\frac{\lambda}{g_{rr}}}
(X^+_6 - X^-_6).
\end{align}
\end{subequations}
We will refer to this inverse transformation when discussing
outer boundary conditions for our numerical simulations in
Sec.~\ref{subsec:Results_KerrSchild}.

%%%%%%%%%%%%%%%%%%%%%%%%%%%%%%%%%%%%%%%%%%%%%%%%%%%%%%%%%%%%%%%
%
%  SECTION 3: DG Method
%
%%%%%%%%%%%%%%%%%%%%%%%%%%%%%%%%%%%%%%%%%%%%%%%%%%%%%%%%%%%%%%%
\section{Discontinuous Galerkin Method}\label{sec:DG}
This section describes the nodal discontinuous 
Galerkin method used to numerically solve 
\eqref{eqn:GBSSNL_spherical}. We adopt a method-of-lines 
strategy, and here describe the relevant semi-discrete scheme
while leaving the temporal dimension continuous. 
To approximate \eqref{eqn:GBSSNL_spherical}, we 
follow the general procedure first introduced in 
Ref.~\cite{Bassi}. Our approach defines local auxiliary 
variables $Q = u'$, and rewrites the spatially second-order 
system \eqref{eqn:GBSSNL_spherical} as the first-order system 
(\ref{eqn:NonconservativeForm}a). Once we use 
(\ref{eqn:NonconservativeForm}b) to eliminate $Q$ from 
(\ref{eqn:NonconservativeForm}a), we recover the {\em primal} 
equations (\ref{eqn:GBSSNL_spherical}). The auxiliary variable 
approach was later generalized and coined the 
{\em local discontinuous Galkerin} (LDG) method in 
Ref.~\cite{CockburnShu}. We may refer to our particular scheme 
as an LDG method, but note that many variations exist in the 
literature. We stress that in LDG methods $Q$ is {\em not} 
evolved and is introduced primarily to assist in the 
construction of a stable scheme.

Equations \eqref{eqn:Aofymatrix} and 
(\ref{eqn:NonconservativeForm}a) imply that the physical flux 
function is
\begin{align}
F(W) = 
\left(
\begin{array}{c}
F_u(W)\\
F_v(W)
\end{array}
\right)
\equiv A(u)W_{v:Q} = 
\left(
\begin{array}{c}
0_{5\times 1}\\
f(W)
\end{array}
\right),\qquad
f =  \left(
\begin{array}{c}
f_{B^r}\\
f_{A_{rr}}\\
f_{K}\\
f_{\Gamma}
\end{array}
\right).
\end{align}
Only the evolution equations for $B^r$, $A_{rr}$, $K$, 
and $\Gamma^r$ give rise to non-zero components in $F$, 
and we have collected these non-zero components into a 
smaller vector $f = F_v$. Inspection 
of \eqref{eqn:GBSSNPrinciplePart} determines these 
components. For example, from
(\ref{eqn:GBSSNPrinciplePart}c) we find
\begin{align}
f_{K} = -\beta^r K + \frac{\chi}{g_{rr}}Q_{\alpha}.
\end{align}

\subsection{Local approximation of the system 
\eqref{eqn:NonconservativeForm}}

Our treatment closely follows \cite{FieldHesthavenLau}, 
but with the equations and notations relevant for this paper.
Our computational domain $\Omega$ is the closed 
$r$-interval $[a,b]$. We cover $\Omega$ with $k_\mathrm{max}>1$ 
non-overlapping intervals $\mathsf{D}^k  = [a^k,b^k]$,
where $a = a^1$, $b = b^{k_\mathrm{max}}$, and $b^{k-1} = a^k$ for 
$k = 2,\cdots,k_\mathrm{max}$. 

On each interval 
$\mathsf{D}^k$, we approximate each component of the system 
vector $W$ by a local interpolating polynomial of 
degree $N$. For example,
\begin{equation}\label{eq:Chiapprox}
\chi^k_h(t,r) = 
\sum_{j=0}^N \chi(t,r_j^k) \ell^k_j(r)
\end{equation}
approximates $\chi(t,r)$.
Throughout this section, approximations
are denoted by a subscript $h$ (see \cite{HEST} for the
notation). For example, $W_h$ and $f_h$
are approximations of $W$ and $f$. Although $Q=u'$,
$Q_h$ and $u_h'$ are not necessarily the same.
In \eqref{eq:Chiapprox} $\ell^k_j(r)$ is the $j$th 
Lagrange polynomial belonging to $\mathsf{D}^k$,
\begin{equation}
\ell^k_j(r) = \prod^{N}_{\stackrel{\scriptstyle 
i = 0}{\scriptstyle i\neq j}}
\frac{r - r^k_i}{r^k_j-r^k_i}.
\end{equation}
Evidently, the polynomial $\chi^k_h$ interpolates $\chi$ at 
$r^k_j$. To define the nodes $r^k_j$, consider the mapping 
from the unit interval $[-1,1]$ to $\mathsf{D}^k$,
\begin{equation}
r^k(u) = a^k + {\textstyle \frac{1}{2}}(1+u)(b^k - a^k),
\end{equation}
and the $N$+1 Legendre-Gauss-Lobatto (LGL) nodes $u_j$. The 
$u_j$ are the roots of the equation
\begin{equation}
(1-u^2)P_N'(u) = 0,
\end{equation}
where $P_N(u)$ is the $N$th degree Legendre polynomial, and the 
physical nodes are simply $r^k_j = r^k(u_j)$.
In vector notation the approximation \eqref{eq:Chiapprox}
takes the form
\begin{equation}
\chi^k_h(t,r) = 
\boldsymbol{\chi}^k(t)^T 
\boldsymbol{\ell}^k(r),
\end{equation}
in terms of the column vectors
\begin{equation}
\boldsymbol{\chi}^k(t) 
= \big[\chi(t,r^k_0),\cdots,
\chi(t,r^k_N)\big]^T,\qquad
\boldsymbol{\ell}^k(r)
= \big[\ell^k_0(r),\cdots,\ell^k_N(r)\big]^T.
\end{equation}

On each open interval $(a^k,b^k) \subset \mathsf{D}^k$ 
and for each component of the equations in 
\eqref{eqn:NonconservativeForm}, we define local residuals 
measuring the extent to which our approximations satisfy
the original continuum system. Dropping the subdomain label 
$k$ on the polynomials and focusing on the
$K$ equation as a representative example, the local 
residual corresponding to (\ref{eqn:GBSSNL_spherical}h) is
\begin{align}\label{eq:residualK}
-(R_K)^k_h  \equiv & 
-\partial_t K_h 
+\left(\beta^r K' \right)_h 
-\left(\frac{\chi Q_{\alpha}'}{g_{rr}}\right)_h 
+\left(\frac{\chi Q_{g_{rr}} Q_{\alpha}}{2g_{rr}^2}\right)_h 
-\left(\frac{\chi Q_{g_{\theta \theta}}
                  Q_{\alpha}}{g_{rr}g_{\theta\theta}
                  }\right)_h 
\nonumber \\
&
+\left(\frac{Q_{\alpha}Q_{\chi}}{2g_{rr}} \right)_h 
+\left(\frac{3\alpha A_{rr}^2}{2g_{rr}^2}\right)_h
+\left(\frac{1}{3}\alpha K^2\right)_h.
\end{align}
Here, for example, the expressions read
\footnote{\label{foot:polyissue}
At this stage the first expression is generically a 
polynomial of degree $2N-1$ and the latter 
is not a polynomial. The conventions adopted in
Eq.~\eqref{eq:hdefforresidual} prove useful while working 
with the residual. However, later on in Sec.~\ref{sec:Together},
to obtain the final form \eqref{eq:semidiscreteK} of the numerical 
approximation corresponding to \eqref{eq:residualK}, 
we will replace nonlinear terms with degree-$N$ polynomials.}
\begin{equation}\label{eq:hdefforresidual}
(\beta^r K')_h = \beta^r_h K_h',\qquad
\left(\frac{Q_\alpha Q_\chi}{2g_{rr}}\right)_h = 
\frac{Q_{\alpha,h} Q_{\chi,h}}{2g_{rr,h}}.
\end{equation}
We similarly construct the remaining eight 
residuals, e.g.~$(R_{g_{rr}})_h$ and $(R_{\Gamma^r})_h$, 
as well as five residuals corresponding to 
(\ref{eqn:NonconservativeForm}b).
For example, one of these remaining five is 
\begin{equation}\label{eqn:RQalpha}
(R_{Q_\alpha})^k_h \equiv -Q_{\alpha,h} + \alpha_h'.
\end{equation}

Let the $k$th inner product be defined as
\begin{equation}
\big(u,v\big)_{\mathsf{D}^k} \equiv \int^{b^k}_{a^k}dr u(r) v(r),
\end{equation}
and consider the expression $(\ell^k_j,(R_K)^k_h)_{\mathsf{D}^k}$.
We call the 
requirement that this inner product vanish $\forall j$ the 
$k$th {\em Galerkin condition}. For each component of the 
system and for each $k$ there is a 
corresponding Galerkin condition, in total 
$9 k_\mathrm{max}(N+1)$ equations for
(\ref{eqn:NonconservativeForm}a) and $5 k_\mathrm{max}(N+1)$ 
for (\ref{eqn:NonconservativeForm}b).
Enforcement of the Galerkin conditions on each $\mathsf{D}^k$ 
will not recover a meaningful global solution, since they 
provide no mechanism for coupling the local 
solutions on the different intervals. Borrowing from the 
finite volume toolbox, we achieve coupling through integration
by parts on $r$ and introduction of the 
{\em numerical flux} $f^*$
at the interface between subdomains. 

In \eqref{eq:residualK} we only 
need to consider $(\beta^r K')_h$ and 
$(\chi Q_{\alpha}'/g_{rr})_h$, 
as the other terms comprise a component of the 
source vector $S_h$. Using integration by parts, we write 
\begin{subequations} \label{eq:IBPK}
\begin{align}
\big(\ell^k_j,(\beta^r K')_h\big)_{\mathsf{D}^k} 
= & -\int_{a^k}^{b^k}dr\left[
     \left(\ell^k_j{} \beta^r_h \right)'{}K_h\right]
  +  \left( \beta^r_h {}K_h \right) \ell^k_j
     \Big|^{b^k}_{a^k}, \\
\big(\ell^k_j,
(\chi Q_{\alpha}'/g_{rr} )_h\big)_{\mathsf{D}^k} 
= & -\int_{a^k}^{b^k}dr\left[
  \left(\ell^k_j{} \frac{\chi_h}{g_{rr,h}} \right)' {}
      Q_{\alpha,h}\right]
+ \left( \frac{\chi_h Q_{\alpha,h}}{g_{rr,h}}\right) 
  \ell^k_j\Big|^{b^k}_{a^k}.
\end{align}
\end{subequations}
In these formulas, we have retained the domain index $k$ 
on $\ell_j^k$, while continuing to suppress it on $K_h$,
$g_{rr,h}$, etc. Moreover, we have suppressed the 
$r$-dependence in all terms on the right-hand side. 
Addition of these formulas along with the definition 
$f_{K,h} = -(\beta^r K )_h + (\chi Q_{\alpha}/g_{rr})_h$
gives 
\begin{align} \label{eqn:IBPK_add}
\big(\ell^k_j,(\beta^r K')_h 
- (\chi Q_{\alpha}'/g_{rr} )_h
\big)_{\mathsf{D}^k} =
-\int_{a^k}^{b^k}dr\left[
\left(\ell^k_j{} \beta^r_h \right)'{}K_h - 
\left(\ell^k_j{} \frac{\chi_h}{g_{rr,h}} \right)'
  {}Q_{\alpha,h} \right]
- f_{K,h} \ell^k_j \Big|^{b^k}_{a^k}.
\end{align}
In lieu of \eqref{eqn:IBPK_add}, we will instead work
with the replacement
\begin{align} \label{eq:IBPK_numericalFlux}
\big(\ell^k_j,(\beta^r K')_h 
- (\chi Q_{\alpha}'/g_{rr} )_h
\big)_{\mathsf{D}^k} \rightarrow
-\int_{a^k}^{b^k}dr\left[
\left(\ell^k_j{} \beta^r_h \right)'{}K_h -
\left(\ell^k_j{} \frac{\chi_h}{g_{rr,h}} \right)'
  {}Q_{\alpha,h} \right]
- f_{K}^* \ell^k_j \Big|^{b^k}_{a^k}.
\end{align}
This replacement features a component $f_{K}^*$ of 
the numerical flux rather than
a component $f_{K,h}$ of the boundary flux. 
The numerical flux is determined by (as yet 
not chosen) functions\footnote{\label{foot:plusminusissue}
In the context of the dG method here, $+$ and $-$ denote 
``exterior" and ``interior", and have no relation to the
$\pm$ using to denote the characteristic fields and speeds
in Table \ref{tab:speeds}. For characteristic fields 
and speeds, $+$ and $-$ mean ``right-moving" and 
``left-moving".}
\begin{align} \label{eq:NumerialFlux_form}
f^* = 
f^*(W^+,W^-),
\end{align}
where, for example, $W^-$ is an interior
boundary value [either $W^k_h(t,a^k)$ or $W^k_h(t,b^k)$] of
the approximation defined on $\mathsf{D}^k$, and
$W^+$ is an exterior
boundary value [either $W^{k-1}_h(t,b^{k-1})$ 
or $W^{k+1}_h(t,a^{k+1})$] of
the approximation defined on either $\mathsf{D}^{k-1}$
or $\mathsf{D}^{k+1}$.
We discuss our choice of numerical flux in the next
subsection. We now employ additional 
integration by parts to write the above
replacement as
\begin{align} \label{eq:IBPK_TwoIBP}
\big(\ell^k_j,(\beta^r K')_h
- (\chi Q_{\alpha}'/g_{rr} )_h
\big)_{\mathsf{D}^k} \rightarrow
\int_{a^k}^{b^k} dr \ell^k_j
\left(\beta^r K' - 
\frac{\chi Q_{\alpha}'}{g_{rr}}\right)_h 
+ \left(f_{K,h} - f_{K}^*
\right)\ell^k_j\Big|^{b^k}_{a^k}.
\end{align}
Rather than the exact $k$th Galerkin condition 
$\big(\ell^k_j,(R_K)^k_h\big)_{\mathsf{D}^k} = 0, \forall j$
for the $K$ component of \eqref{eqn:NonconservativeForm}
on $\mathsf{D}^k$, we will instead strive to enforce
\begin{align} \label{eqn:GalerkinK}
\big(\ell^k_j,(R_K)^k_h\big)_{\mathsf{D}^k} =
\left(f_{K,h} -  f_{K}^*\right)\ell^k_j
\Big|^{b^k}_{a^k},\quad \forall j
\end{align}
although our treatment of nonlinear 
terms will lead to a slight modification of
these equations (we return to this issue shortly).
The other components of (\ref{eqn:NonconservativeForm}a) 
are treated similarly, as are the components of 
(\ref{eqn:NonconservativeForm}b).
Recall that, for example, $Q_{\alpha} = \alpha'$.
Formally using the same dG method to solve for $Q_{\alpha}$, 
we arrive at the replacement
\begin{align}
\big(\ell^k_j,(R_{Q_\alpha})^k_h\big)_{\mathsf{D}^k} 
\rightarrow & \int_{a^k}^{b^k} dr \ell^k_j \left(
-Q_{\alpha,h} + \alpha_h '\right)
- \left(\alpha_h - \alpha^*\right)\ell^k_j
\Big|^{b^k}_{a^k}, 
\end{align}
which again features a component $\alpha^*$ of the 
numerical flux. The auxiliary variables are constructed 
and used at each stage of temporal integration. 
We then have
\begin{equation}
\big(\ell^k_j,(R_{Q_\alpha})^k_h\big)_{\mathsf{D}^k}
= \left(\alpha_h - \alpha^*\right)\ell^k_j
\Big|^{b^k}_{a^k}, \quad \forall j
\end{equation}
as the corresponding enforced $k$th Galerkin condition.

\subsection{Numerical Flux}
\label{sec:NFsec}
To further complete our dG scheme we must specify 
functional forms for the components of the numerical 
flux introduced in the previous section. We distinguish 
between the physical fluxes (components of $f$) and 
the auxiliary fluxes (components of $u$) arising from 
the definition of the auxiliary variables. These 
choices are not independent as the resulting scheme 
must be stable and consistent. Our choice follows 
\cite{Brezzi_FluxChoice} which considered diffusion 
problems. Additional analysis of this flux choice 
appears in \cite{Arnold_dG,HEST}.

Let us first consider the numerical fluxes corresponding 
to the physical fluxes and of the form 
(\ref{eq:NumerialFlux_form}). The numerical flux vector 
is a function of the system and auxiliary variables 
interior and exterior to a subdomain. A common choice 
for $f^*$ is 
\begin{align}\label{eqn:f_fluxBSSN}
f^*= \lcurlyjump           f_h \rcurlyjump 
   + \frac{\tau}{2} \ljump v_h \rjump,\qquad
K\text{-component of $f^*$: }
f^*_K = \lcurlyjump           f_{K,h} \rcurlyjump
   + \frac{\tau}{2} \ljump K_h \rjump,
\end{align}
where, as an example, we have also shown the 
component of $f^*$ corresponding to the analysis 
above. Respectively, the average and jump 
across the interface are
\begin{align}
\lcurlyjump f_h  \rcurlyjump = 
       \frac{1}{2}\left(f^+ + f^-\right),
\quad  \ljump v_h \rjump = 
       \mathbf{n}^- v^- + \mathbf{n}^+v^+.
\end{align}
Here $\tau$ is a position dependent penalty parameter 
(fixed below) and $\mathbf{n}^-(\mathbf{n}^+)$ is the local 
outward pointing normal to the interior (exterior) 
subdomain. The role of $\tau$ is to ``penalize" 
(i.~e.~yield a negative contribution to the $L_2$ 
energy norm) jumps across an interface. An 
appropriate choice of $\tau$ will ensure stability,
and we now provide some motivation for the choice 
\eqref{eqn:PenaltyScaling} of $\tau$ we make below.

Were we treating the fully first-order system 
\eqref{eqn:firstordersys_abs}, the local 
Lax-Friedrichs flux would often 
be a preferred choice due to its simplicity 
\cite{HEST}. In this case, the constant 
$\omega$ in the numerical flux formula
$\mathcal{F}^* = 
\lcurlyjump \mathcal{F}_h\rcurlyjump
+ \frac{1}{2}\omega \ljump W_h \rjump
$
obeys 
$\omega \geq \mathrm{max} \big|\mu(\nabla_W\mathcal{F}(W)
) \big|$. Here, $\mathcal{F}(W) = \mathcal{A}(u)W$, the 
notation $\mu(\cdot)$ indicates the
spectral radius of the matrix within, and the 
max is taken over interior $W^-$ and exterior 
$W^+$ states. Motivated by \eqref{eqn:partofpp}, we
adopt a similar but simpler prescription, 
substituting the field gradient 
\begin{equation}
\nabla_{W_{v:Q}} \tilde{A}(u) W_{v:Q} = \tilde{A}(u)
\end{equation}
for $\nabla_W\mathcal{F}(W)$. 
Precisely, we assume the scaling
\begin{equation} \label{eqn:PenaltyScaling}
\tau(b^k)=\tau(a^{k+1})=\tau^{k+1/2}
           \equiv 
           C \cdot \mathrm{max}
           \big|\mu 
           \big(\tilde{A}(u)
           \big)
           \big|,
\end{equation}
where $C = O(1)$ is a constant chosen for stability. 
Larger values of $C$ will result in schemes with better stability 
properties, whereas too large a value will impact the CFL 
condition.
At the interface point $\mathsf{I}^{k+1/2} \equiv 
\mathsf{D}^k \cap \mathsf{D}^{k+1}$, the vector $u_h$
has two representations: $u^{-}$ at $b^k$ and $u^+$
at $a^{k+1}$. The max in \eqref{eqn:PenaltyScaling} is taken 
over the corresponding two sets of field speeds. More precisely, 
the speeds in Table \ref{tab:speeds} are computed for both 
$u^{-}$ and $u^+$, and the maximum taken over all resulting
speeds. For the auxiliary variables, a penalized central 
flux is used The definition with one representative 
component is
%\begin{align}\label{eqn:alpha_fluxBSSN}
%u^* = \lcurlyjump u_h \rcurlyjump
%- \frac{\tau_u}{2} \ljump u_h \rjump,
%\qquad \alpha\text{-component of $u^*$: }
%\alpha^* = \lcurlyjump\alpha_h \rcurlyjump
%- \frac{\tau_{\alpha}}{2} \ljump \alpha_h \rjump,
%\end{align}
\begin{align}\label{eqn:alpha_fluxBSSN}
u^* = \lcurlyjump u_h \rcurlyjump
- \frac{1}{2} \ljump u_h \rjump,
\qquad \alpha\text{-component of $u^*$: }
\alpha^* = \lcurlyjump\alpha_h \rcurlyjump
- \frac{1}{2} \ljump \alpha_h \rjump,
\end{align}
with similar expressions for the remaining components.
%For auxiliary variables we will always view $\tau_u$
%as the same fixed scalar for all components of $u$, 
%and set $\tau_u = 1 = \tau_\alpha$, etc.

We stress the following point. Since the interior coupling
between subdomains is achieved through the numerical flux forms
\eqref{eqn:PenaltyScaling} and \eqref{eqn:alpha_fluxBSSN}, 
the inverse transformation \eqref{eqn:fundvarsFROMcharvars} expressing
the fundamental fields in terms of the characteristic fields is not
required to achieve this coupling. On the other hand, imposition of 
physical boundary conditions may still rely on 
\eqref{eqn:fundvarsFROMcharvars}, since this transformation allows
one to fix only incoming characteristic modes. 

\subsection{Nodal form of the semi-discrete equations}
\label{sec:Together}
Let us introduce the $k$th
{\em mass} and {\em stiffness} matrices,
\begin{equation}
M_{ij}^k=\int_{a^k}^{b^k}dr \ell_i^k(r) \ell_j^k (r),\qquad
S_{ij}^k =\int_{a^k}^{b^k}dr \ell_i^k(r)\ell_j^k{}'(r).
\end{equation}
These matrices belong to $\mathsf{D}^k$, and the corresponding
matrices defined on the reference interval $[-1,1]$ are 
\begin{equation}
\bar{M}_{ij} = \int_{-1}^1 du \ell_i(u) \ell_j(u),\qquad
\bar{S}_{ij} = \int_{-1}^1 du \ell_i(u) \ell_j'(u),
\end{equation}
where $\ell_j(u)$ is the $j$th Lagrange polynomial determined
by the LGL nodes $u_j$ on $[-1,1]$. These matrices are related 
by $M^k_{ij} = {\textstyle \frac{1}{2}}(b^k-a^k) \bar{M}_{ij}$ 
and $S^k_{ij} = \bar{S}_{ij}$, whence only the reference matrices
require computation and storage. 

We will use the matrices $M^k$ and $S^k$ in obtaining an ODE 
system from \eqref{eq:residualK} and \eqref{eqn:GalerkinK}.
Towards this end, we first approximate the nonlinear terms
(products and quotients) in \eqref{eq:residualK} by degree-$N$
interpolating polynomials. Such approximations are achieved through
pointwise representations. For example, 
$(Q_\alpha Q_\chi/g_{rr})_h$ appears 
in \eqref{eq:residualK}, 
and is expressed in the following way:
[cf.~footnote \ref{foot:polyissue}]
\begin{equation}\label{eq:productexpansion}
\left(
\frac{Q_\alpha Q_\chi}{g_{rr}}\right)_h(t,r) 
= \left(\frac{Q_{\alpha,h} Q_{\chi,h}}{g_{rr,h}}\right)
(t,r)
\rightarrow
\sum_{j=0}^N \frac{Q_{\alpha,h}(t,r^k_j)
                   Q_{\chi,h}(t,r^k_j)}{
                   g_{rr,h}(t,r^k_j)}
                   \ell^k_j(r).
\end{equation}
Note that the expressions on the right and left are not
equivalent due to aliasing error \cite{HEST_GOT}. Our vector notation
for this replacement will be 
\begin{equation}
\left(\frac{Q_\alpha Q_\chi}{g_{rr}}\right)_h(t,r) \rightarrow
\left(\frac{\boldsymbol{Q}_\alpha
            \boldsymbol{Q}_\chi}{
            \boldsymbol{g}_{rr}}\right)(t)^T
\boldsymbol{\ell}^k(r).
\end{equation}
Operations among bold variables are always performed pointwise.
Making similar replacements for all terms in \eqref{eq:residualK},
and then carrying out the integrations in \eqref{eqn:GalerkinK},
which bring in $M^k$ and $S^k$, we arrive at
\begin{align}\label{eq:semidiscreteK}
   \partial_t \boldsymbol{K} & = 
   \boldsymbol{\beta}^r D \boldsymbol{K} 
  -\frac{\boldsymbol{\chi} D 
               \boldsymbol{Q}_{\alpha}}{
               \boldsymbol{g}_{rr}}
  +\frac{1}{2}
   \frac{\boldsymbol{\chi} \boldsymbol{Q}_{g_{rr}} 
               \boldsymbol{Q}_{\alpha}}{
               \boldsymbol{g}_{rr}^2}
  -\frac{\boldsymbol{\chi}
               \boldsymbol{Q}_{g_{\theta \theta}} 
               \boldsymbol{Q}_{\alpha}}{
               \boldsymbol{g}_{rr}
               \boldsymbol{g}_{\theta \theta}} \nonumber \\
& +\frac{1}{2}
   \frac{\boldsymbol{Q}_{\alpha} 
               \boldsymbol{Q}_{\chi}}{
               \boldsymbol{g}_{rr}}
  +\frac{3}{2}
   \frac{\boldsymbol{\alpha} 
               \boldsymbol{A}_{rr}^2}{
               \boldsymbol{g}_{rr}^2}
  +\frac{1}{3}
   \boldsymbol{\alpha} 
         \boldsymbol{K}^2
  + M^{-1}\boldsymbol{\ell}^k
    \left(f_{K,h} - f_{K}^*\right)
    \Big|^{b^k}_{a^k},
\end{align}
where we have again suppressed the superscript $k$ on
all terms except $\boldsymbol{\ell}^k(r)$, and the subscript
$h$ is dropped on all boldfaced variables. 
As described in \cite{HEST}, the spectral collocation 
derivative matrix
\begin{equation}
(D^k)_{ij} = \left.
\frac{d\ell^k_j}{dr}\right|_{r = r^k_i}
\end{equation}
can also be expressed as $D^k = (M^k)^{-1} S^k$,
which appears above. 
Eight other semi-discrete
evolution equations are similarly obtained, with nine in total 
(one for each component of $W_{u:v}$). Additionally, we have
\begin{equation}
\boldsymbol{Q}_\alpha = D\boldsymbol{\alpha} + M^{-1}
\boldsymbol{\ell}^k 
(\boldsymbol{\alpha}^* - \boldsymbol{\alpha}_h)\Big|^{b_k}_{a^k}. 
\end{equation}
as one of the auxiliary equations, with five in total (one for each 
component of $Q = W_{Q:Q}$).

\subsection{Filtering}
Like other nodal (pseudospectral) methods, our scheme may suffer 
from instabilities driven by aliasing error \cite{HEST_GOT}.
{\em Filtering} is a simple yet robust remedy. 
To filter a solution component, such
as $\chi$, we use the modal (as opposed to nodal) representation of
the solution:
\begin{equation}\label{eq:modal_rep}
\chi^k_h(t,r) = 
\sum_{j=0}^N \chi(t,r_j^k) \ell^k_j(r) = 
\sum_{j=0}^N \hat{\chi}^k_j(t) P_j(r),
\end{equation}
where $P_j(r)$ is the $j$th Legendre polynomial. 
Let $\eta_j = j/N$, and define the filter function 
\begin{equation} \label{eqn:filter}
\sigma(\eta_j) = 
\left\{
\begin{array}{lcl}
1 & & \text{for } 0 \leq \eta_j \leq N_c/N\\
\exp
\Big(-\epsilon \Big(
               \frac{\eta_j-N_c/N}{1-N_c/N}
               \Big)^{2s}
\Big) & & \text{for } N_c/N \leq \eta_j \leq 1.
\end{array}\right.
\end{equation}
At each timestep we modify our solution component according to 
\begin{equation}
\chi^k_h \rightarrow 
\left(\chi^k_h\right)^{\mathrm{filtered}} = 
\sum_{j=0}^N \sigma(\eta_j) \hat{\chi}^k_j(t) P_j(r) .
\end{equation}
Evidently, the modification only affects the top $N-N_c$ modes,
and is sufficient to control the type of 
weak instability driven by aliasing \cite{HEST}. 
The numerical parameters $N_c$ and $\epsilon$ are problem 
dependent. For our simulations, we have taken $\epsilon \simeq 
-\mathrm{log}(\varepsilon_\mathrm{mach}) = 36$, where 
$\varepsilon_\mathrm{mach}$ is machine accuracy in double precision.

\subsection{Model system}
\label{sec:ToyPDE}
To better illustrate the basic properties of our method, we
consider a toy model. Namely, the following spatially second-order 
system:
\begin{subequations} \label{eqn:TOY}
\begin{align}
\partial_t u & =  u' + av - u^3 + g(t,x) \\ 
\partial_t v & =  u'' + v'- (u+v)(u')^2 + v^2u^2 + h(t,x),
\end{align}
\end{subequations}
where $a \geq 1$ is constant and $g$ and $h$ are analytic source terms 
to be specified. In contrast to \eqref{eqn:vectors_uvQ},
here $u$, $v$, and $Q = u'$ are scalars 
rather than vectors. System (\ref{eqn:TOY}) admits a first-order 
reduction in which $u'$ is defined as an extra variable. Since
this first-order reduction is strongly hyperbolic, the spatially second-order 
system (\ref{eqn:TOY}) is also strongly hyperbolic by one of the 
definitions considered in \cite{Gundlach_2006}. The characteristic 
fields $X^\pm$ and speeds $\mu^\pm$ are
\begin{equation}\label{eqn:toyXandLambda}
X^+       = \sqrt{a} v - u',\quad 
\mu^+ = \sqrt{a} - 1;
\qquad 
X^-       = \sqrt{a} v + u',\quad 
\mu^- = -(\sqrt{a} + 1).
\end{equation}
To construct a local dG scheme for this system, we 
first rewrite it as 
\begin{subequations}
\begin{align}
\partial_t u & =  Q + av  - u^3 + g(t,x) \\ 
\partial_t v & =  Q' + v'- (u+v)Q^2 + v^2u^2 + h(t,x) \\
Q            & = u'.
\end{align}
\end{subequations}
Evidently, $f=-(Q+v)$ is the $v$-component of the 
physical flux vector
\begin{equation}
F(v,Q) \equiv \left(\begin{array}{c}F_u\\F_v\end{array}\right) = 
\left(\begin{array}{c}0\\f\end{array}\right). 
\end{equation}
Note that $F$ has the same structure as $(u,v)^T$.
Borrowing from the presentation for the BSSN system, we write 
the analogous semidiscrete scheme on each subdomain $\mathsf{D}^k$ 
for the model system:
\begin{subequations} \label{eqn:TOY_SD}
\begin{align}
\partial_t \boldsymbol{u} & = 
\boldsymbol{Q} + a\boldsymbol{v} 
- \boldsymbol{u}^3 + \boldsymbol{g}(t)\\ 
\partial_t \boldsymbol{v} & = D\boldsymbol{Q} + D\boldsymbol{v} 
 - (\boldsymbol{u}+\boldsymbol{v})\boldsymbol{Q}^2 
 + \boldsymbol{v}^2\boldsymbol{u}^2 + \boldsymbol{h}(t)
+ M^{-1}\boldsymbol{\ell}^k  (f_h - f^*) \Big|^{b^k}_{a^k}, \\
\boldsymbol{Q} & = D \boldsymbol{u} + M^{-1}\boldsymbol{\ell}^k
(u^* - u_h)\Big|^{b^k}_{a^k}.
\end{align}
\end{subequations}
Here, we have suppressed the subinterval label $k$ from all 
variables except for the vector $\boldsymbol{\ell}^k$ of Lagrange 
polynomial values. Moreover, following the guidelines discussed
above, the numerical fluxes are given by
\begin{equation}\label{eqn:toyfluxes}
f^* = \lcurlyjump f_h \rcurlyjump 
      + \frac{1+\sqrt{a}}{2} \ljump v_h \rjump,
\qquad
u^* = \lcurlyjump u_h \rcurlyjump 
      - \frac{1}{2} \ljump u_h \rjump.
\end{equation}
Appendix \ref{WPToy} analyzes the stability of our scheme, 
for a more general numerical flux choice, as
applied to (\ref{eqn:TOY}) with the nonlinear and source terms
dropped.

%%%%%%%%%%%%%%%%%%%%%%%%%%%%%%%%%%%%%%%%%%%%%%%%%%%%%%%%%%%%%%%
%
%  SECTION 4: RESULTS
%
%%%%%%%%%%%%%%%%%%%%%%%%%%%%%%%%%%%%%%%%%%%%%%%%%%%%%%%%%%%%%%%
\section{Results from numerical simulations}\label{sec:NumSims}

This section presents results found by numerically solving both 
the model system (\ref{eqn:TOY}) and
BSSN system \eqref{eqn:GBSSNL_spherical} 
with the dG scheme presented in Sec.~\ref{sec:DG}.  
%%%%%%%%%%%%%%%%%%%%%%%%%%%%%%%%%%%%%%%%%%%%%%%%%%%%%%%%%%
% Figure of Test problem
\begin{figure}
\centering
\includegraphics[height=4.75in]{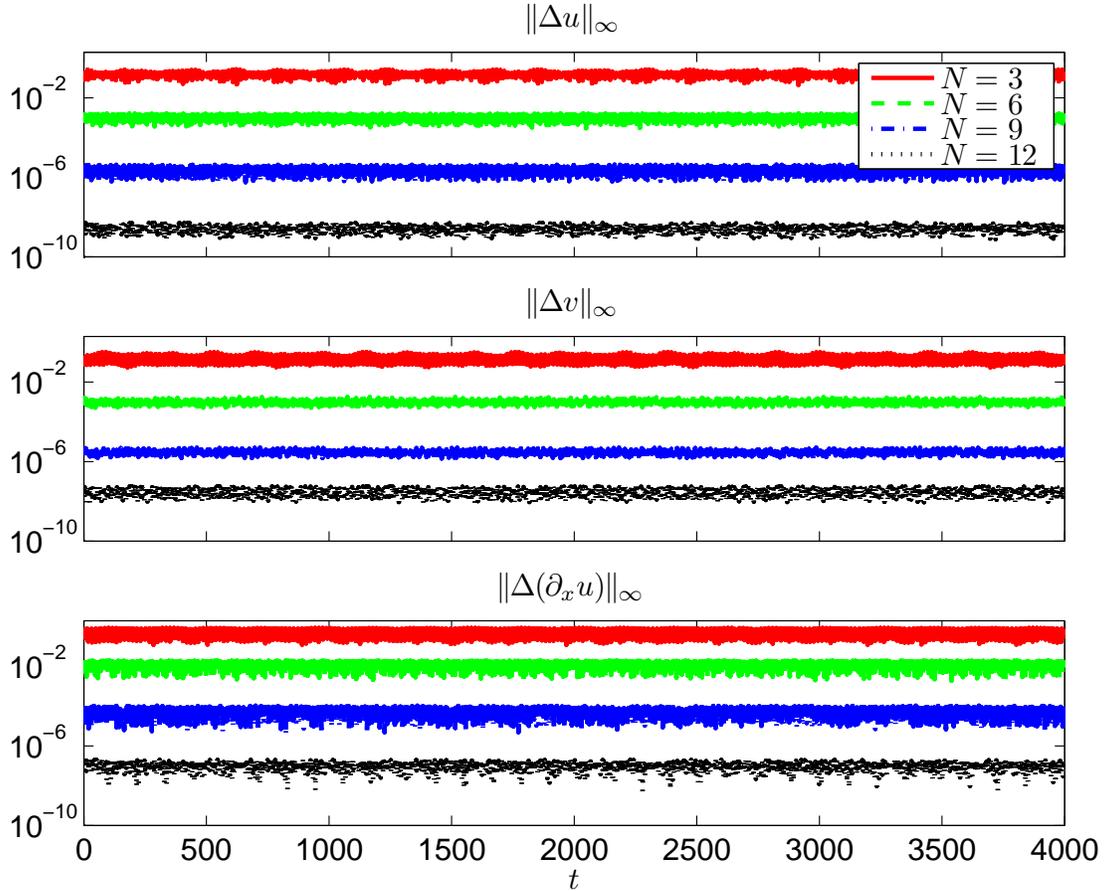}
\caption{{\sc Spectral convergence of fields for model PDE.}
Respectively, for $N = 3,6,9,12$, a timestep of
$\Delta t = 0.0578, 0.0178,0.0084,0.0049$ has been chosen
for stability and accuracy. In the title headings, for example,
$\Delta u \equiv u_\mathrm{numer} - u_\mathrm{exact}$.
}
\label{fig:TESTSpec}
\end{figure}
%%%%%%%%%%%%%%%%%%%%%%%%%%%%%%%%%%%%%%%%%%%%%%%%%%%%%%%%%%

\subsection{Simulations of the model system}\label{subsec:toy_simulations}
\begin{figure}
\centering
\includegraphics[height=3.75in]{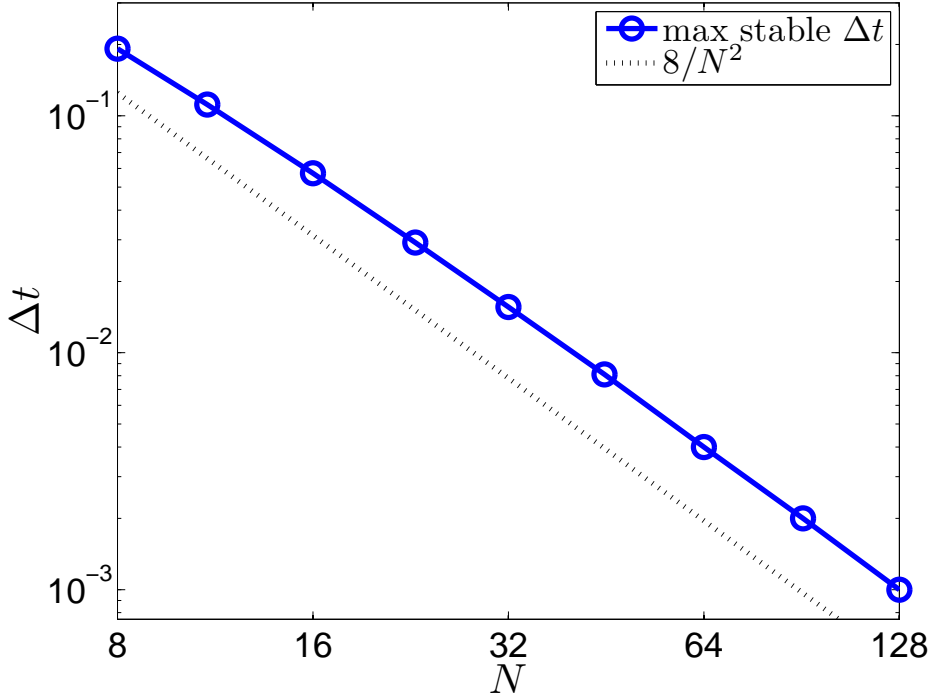}
\caption{{\sc Scaling of maximum stable 
$\Delta t$ with $N$ for model PDE.}
}
\label{fig:TEST_dtScaling}
\end{figure}

The semi-discrete scheme (\ref{eqn:TOY_SD}) has been integrated with 
the classical fourth-order Runge-Kutta method. When integrating this 
system, we have first constructed $\boldsymbol{Q}$ at each Runge-Kutta 
stage, and then substituted into the evolution equations 
(\ref{eqn:TOY_SD}a,b) for $\boldsymbol{u}$ and $\boldsymbol{v}$. The problem has been 
solved on a computational domain $[0,4\pi]$ comprised of two subdomains 
with a timestep chosen small enough for stability. 
The initial data has been taken
from the following exact solution to (\ref{eqn:TOY}): 
\begin{subequations} \label{eqn:ToyParticular}
\begin{align}
u'_\mathrm{exact}(t,x) & =
\frac{1}{2}\big[\sin(x - \mu^- t)
               -\sin(x - \mu^+ t)
           \big] \\
v_\mathrm{exact}(t,x)  & =
\frac{1}{2\sqrt{a}}\big[\sin(x - \mu^- t)
                       +\sin(x - \mu^+ t)
                   \big] \\
g(t,x) & = u_\mathrm{exact}^3 \\
h(t,x) & = (u_\mathrm{exact} + v_\mathrm{exact})
      (u_\mathrm{exact}')^2 -
       v_\mathrm{exact}^2 u_\mathrm{exact}^2,
\end{align}
\end{subequations}
where the speeds $\mu^\pm$ are found in 
\eqref{eqn:toyXandLambda}.
Specification of the boundary condition at a physical
endpoint amounts to choosing the external 
state for at the endpoint. We have considered 
two possibilities: (i) the analytic
state $(Q^+,v^+) = (Q_\mathrm{exact},v_\mathrm{exact})$
and (ii) an upwind state. For example, at $x= 4\pi$ the
upwind state is\footnote{We remind the reader that,
unfortunately, the $\pm$ on $X^\pm$ means something 
different than the $\pm$ indicating exterior/interior 
dG states [cf.~footnote \ref{foot:plusminusissue}].}
\begin{equation}\label{eqn:toy_upwind_state}
Q^+ = Q_\mathrm{upwind} = 
\frac{1}{2}
\big[(X^-)_\mathrm{exact}
    -(X^+)_\mathrm{numer}\big],
\quad
v^+ = v_\mathrm{upwind} = 
\frac{1}{2\sqrt{a}}
\big[(X^-)_\mathrm{exact}
    +(X^+)_\mathrm{numer}\big].
\end{equation}
Either choice of $(Q^+,v^+)$ leads to similar results, 
and the plots here correspond to the analytic state. 
Figure \ref{fig:TESTSpec} clearly shows spectral convergence 
with increasing polynomial order $N$ across all fields 
for the case $a=2$. Other values of $a$, 
including $a=1$ for which $X^+$ is a static characteristic 
field, have also been considered with similar results.
Appendix \ref{WPToy} demonstrates that our proposed scheme
for the system \eqref{eqn:TOY_SD} with nonlinear and source
terms dropped is stable in a semi-discrete sense. Nevertheless, the
fully discrete scheme, obtained via temporal discretization 
by the fourth-order Runge-Kutta method, is still subject to
the standard absolute stability requirement. Namely, if 
$\mu_h$ is any eigenvalue corresponding to the (linearized)
discrete spatial operator, then a necessary condition for 
stability is that $\mu_h \Delta t$ lies in 
{\em absolute stability region} for 
Runge-Kutta 4. We here show empirically 
that the associated timestep restriction scales like $N^{-2}$,
i.e.~$\Delta t = O(N^{-2})$ for stability. We note that 
such scaling is welcome in light of the second-order 
spatial operators which appear in the system, and suggest
a possible worse scaling like $N^{-4}$. Fig.~\ref{fig:TEST_dtScaling} 
plots the maximum stable timestep for a range of $N$, 
demonstrating the $N^{-2}$ scaling,
in line with behavior known from analysis of 
first-order systems \cite{HEST}. This scaling also holds 
for the BSSN system.
%%%%%%%%%%%%%%%%%%%%%%%%%%%%%%%%%%%%%%%%%%%%%%%%%%%%%%%%%%%%%%%%%%%%%%%%%%%%%%%%
% Figure of Test problem
%%%%%%%%%%%%%%%%%%%%%%%%%%%%%%%%%%%%%%%%%%%%%%%%%%%%%%%%%%%%%%%%%%%%%%%%%%%%%%%%

\subsection{Simulations of the BSSN system}\label{subsec:Results_KerrSchild}
This subsection documents results for simulations of the unit-mass-parameter
($M = 1$) Schwarzschild solution (\ref{eq:BSSN_KS}) expressed in terms of 
ingoing Kerr-Schild coordinates. Since the solution is stationary, temporal
integration of the semi-discrete scheme has been carried out with the
forward Euler method which the dissipation in our method allows. 
The $r$-coordinate domain $[0.4, 3.4]$ has been split into 3 equally spaced
subdomains, and we have set $\eta = 10$, $\lambda = 0.1$, and 
$C = 2$ [cf.~Eq.~\eqref{eqn:PenaltyScaling}]. For all simulations $\Delta t$ 
has been chosen for stability. With the chosen $\lambda$, the inner 
physical boundary $r_\mathrm{min} = 0.4$ is an excision surface. At each timestep
we have applied an (order $2s=20$) exponential filter on the top two-thirds
of the modal coefficient set for all fields {\em except} for $g_{rr}$ and 
$g_{\theta \theta}$. {\em For stability, we have empirically
observed that $g_{rr}$ and $g_{\theta \theta}$ must not be filtered.}
A detailed understanding of this is still lacking.

Issues related to physical boundary conditions are similar to the one
encountered in Sec.~\ref{subsec:toy_simulations} for the model problem. 
Similar to before, we have retained 
Eqs.~(\ref{eqn:f_fluxBSSN},\ref{eqn:alpha_fluxBSSN}) as the choice 
of numerical flux even at the endpoints. 
Therefore, at an endpoint the specification of
the boundary condition amounts to the choice $W^+$ of external state.
We have typically chosen the inner boundary of the radial domain as 
an excision boundary, and in this case $W^+ = W^-$ is enforced at 
the inner physical boundary. At the outer physical boundary,
for $W^+$ we have again considered two choices: (i) $W_\mathrm{exact}$ 
and (ii) $W_\mathrm{upwind}$. To enforce choice (ii) the inverse
transformation \eqref{eqn:fundvarsFROMcharvars} must be used with
incoming characteristic fields fixed to their exact values, similar to
\eqref{eqn:toy_upwind_state}. We have tried various versions of choice
(ii), and in all cases the resulting simulations have been unstable.
We therefore present results corresponding to choice (i). 
Although the choice of an analytical external state $W_\mathrm{exact}$ 
at the outer boundary is stable for our problem, such a boundary 
condition is unlikely to generalize to more complicated scenarios 
involving dynamical fields. Indeed, the issue
of outer boundary conditions for the BSSN system is an active area 
of research, with a proper treatment requiring fixation of incoming 
radiation, control of the constraints, and specification of gauge
(see Ref.~\cite{NunezSarbach2009} for a recent analysis).
%%%%%% KS initial data - conformally flat
\begin{figure}
\centering
\includegraphics[height=4.75in]{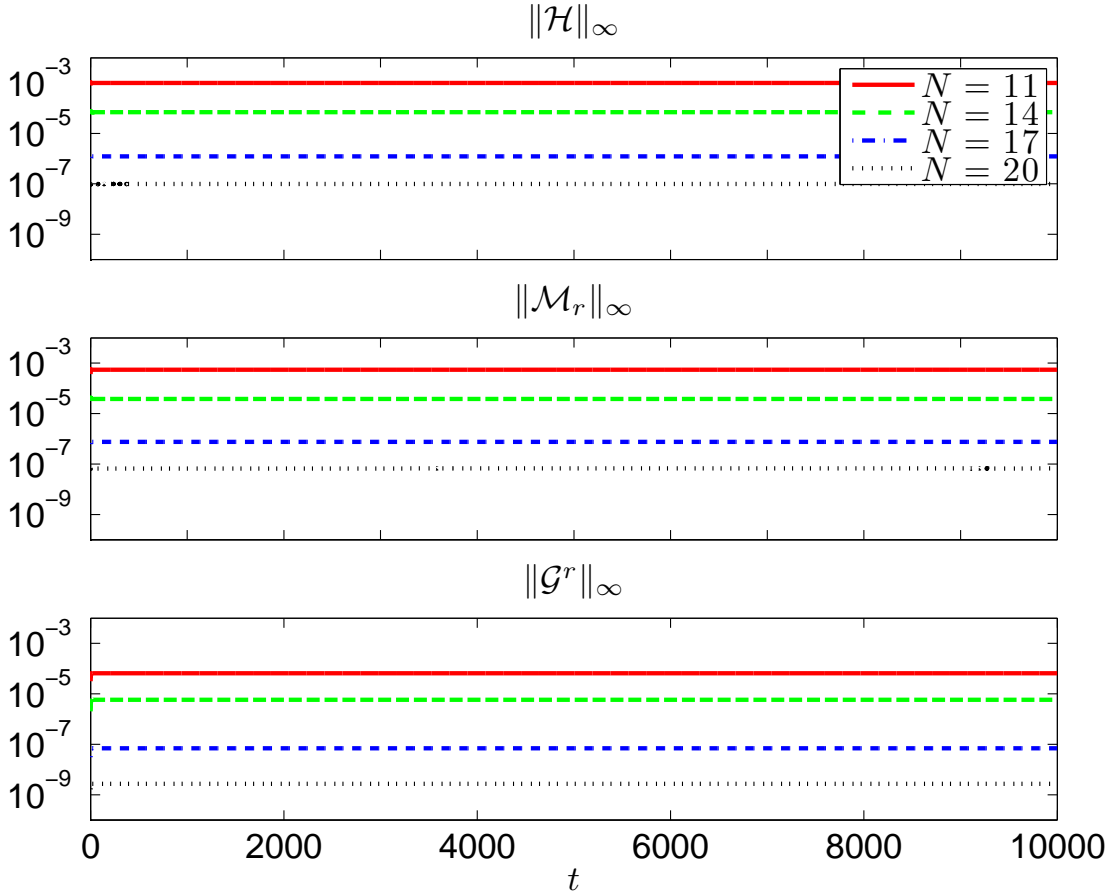}
\caption{{\sc Spectral convergence of constraint violations for $M=1$
Kerr-Schild initial data.}
Respectively, for $N = 11,14,17,19$, a timestep of
$\Delta t \simeq 0.0041,0.0026,0.0018,0.0013$ has been chosen
for stability and accuracy.}
\label{fig:M1_BHFull_constraints_conform}
\end{figure}
\begin{figure}
\centering
\includegraphics[height=4.75in]{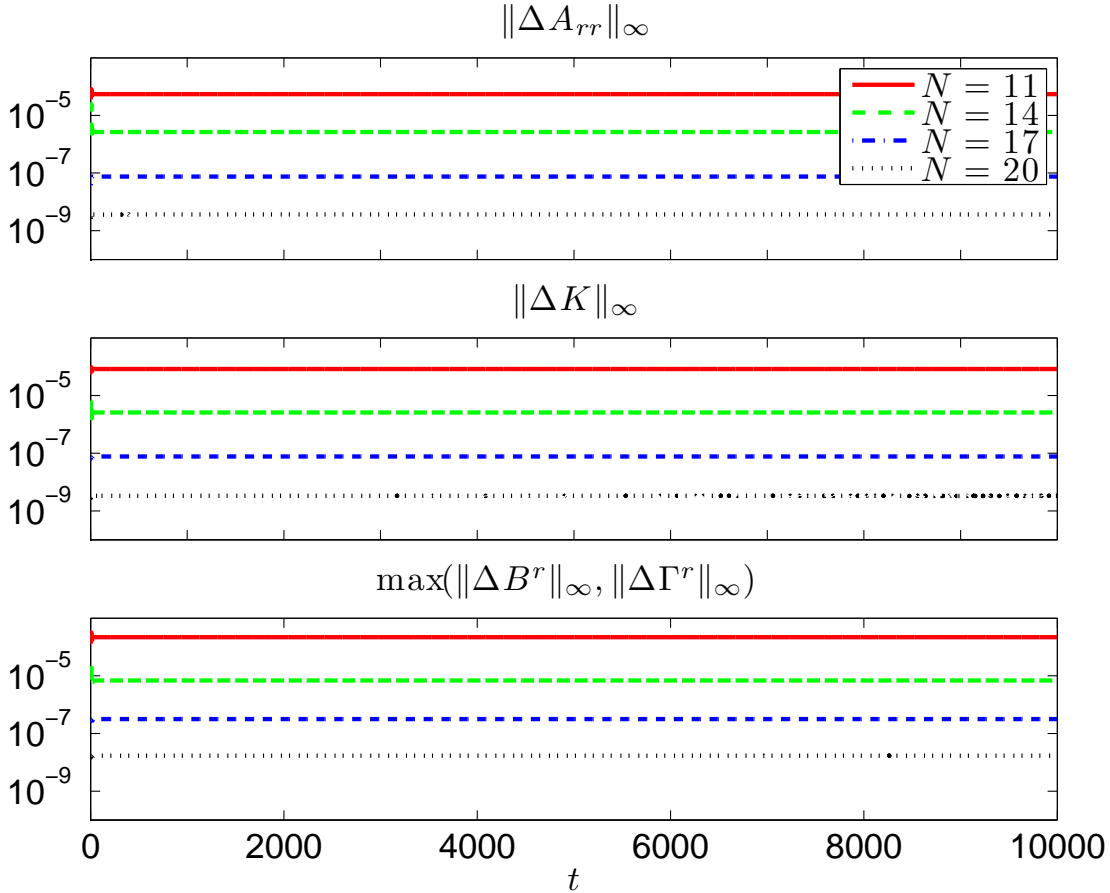}
\caption{{\sc Spectral convergence of solution for $M=1$
Kerr-Schild initial data.} Timestep choices are described in
the caption for Fig.~\ref{fig:M1_BHFull_constraints_conform}.
In the title headings, for example,
$\Delta A_{rr} \equiv (A_{rr})_\mathrm{numer}
- (A_{rr})_\mathrm{exact}$.
}
\label{fig:M1_BHFull_Fields1_conform}
\end{figure}
\begin{figure}
\centering
\includegraphics[height=4.75in]{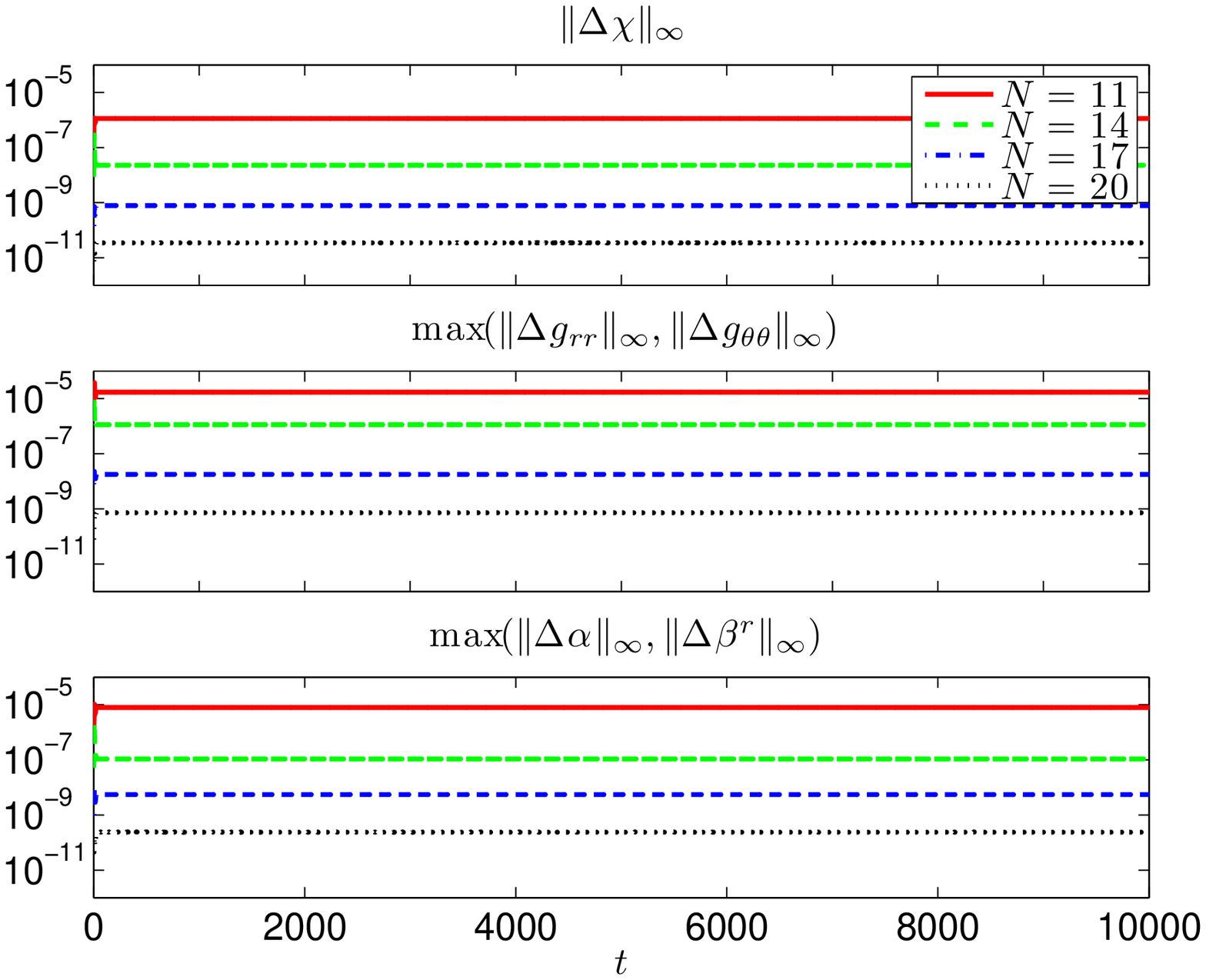}
\caption{{\sc Spectral convergence of solution violations for $M=1$
Kerr-Schild initial data.} See the caption of
Fig.~\ref{fig:M1_BHFull_Fields1_conform} for details.}
\label{fig:M1_BHFull_Fields2_conform}
\end{figure}
%%%%%%%%%%%%%%%%%%%%%%%%%%%%%%%%%%%%%%%%

For BSSN simulations, our main diagnostic is to monitor the 
Hamiltonian, momentum, and conformal connection constraints.
Figure \ref{fig:M1_BHFull_constraints_conform}
depicts long-time histories of constraint violations, 
whereas Figs.~\ref{fig:M1_BHFull_Fields1_conform} and 
\ref{fig:M1_BHFull_Fields2_conform} depict long-time 
error histories for the individual BSSN field components. 
From the middle plot in Fig.~\ref{fig:M1_BHFull_Fields2_conform},
we infer that, up to the indicated numerical error, the factor 
$g/\sin^4\theta = g_{rr} (g_{\theta\theta})^2$ 
remains at its initial fixed profile $r^4$ throughout the evolution.
These figures indicate that the proposed scheme is stable 
for long times, and exhibits spectral converge with 
increased polynomial order $N$. Similar results are recovered 
from $M=0$ Minkowski initial data. The stability documented 
in these plots does not appear to rely on inordinate 
parameter tuning. For example, with the fixed parameters 
described above, we obtain similar plots if we 
individually vary (i) $r_\mathrm{min}$ over 
$\{0.325, 0.35, 0.4, 0.475\}$ (values still corresponding 
to an excision surface for the given choice of $\lambda$), 
(ii) $\eta$ over $\{1, 3, 7, 10\}$, 
(iii) $s$ over $\{8,9,10\}$. With the polynomial order 
$N$ ranging over $\{23, 26, 29, 31\}$, both stability and 
qualitatively similar exponential convergence is achieved 
with a single subdomain. Likewise, adoption of a larger 
coordinate domain with more subdomains does not significantly 
impact our results. However, for much larger $r_\mathrm{max}$ 
stability requires a smaller time step or a time stepper better
suited for wave problems (e.g.~Runge Kutta 4).
Finally, we have considered the addition of random 
noise to all field components at the initial time. Precisely, 
with the system component $\chi$ as an example, we have set
\begin{equation}
\boldsymbol{\chi}(0) \mapsto \boldsymbol{\chi}(0)
+ \delta \boldsymbol{\chi}(0),
\end{equation}
where each component (nodal value) of 
$\delta\boldsymbol{\chi}(0)$ is
$10^{-5}$ times a random variable drawn from a standard
normal distribution. Such perturbed initial data also
gives rise to stable evolutions.

%%%%%%%%%%%%%%%%%%%%%%%%%%%%%%%%%%%%%%%%%%%%%%%%%%%%%%%%%%%%%%%
%
%  SECTION 5: FINAL REMARKS
%
%%%%%%%%%%%%%%%%%%%%%%%%%%%%%%%%%%%%%%%%%%%%%%%%%%%%%%%%%%%%%%%

\section{Conclusion}\label{sec:conclusion}
We have introduced a discontinuous Galerkin method for solving 
the spherically reduced BSSN system with second-order spatial
operators. Our scheme shares similarities with other
discontinuous Galerkin methods that use local auxiliary 
variables to handle high-order spatial derivatives 
\cite{Bassi,
CockburnShu,
ShuSchro,
ShuKvd,
Arnold_dG,
Brezzi_FluxChoice,
HEST},
and which have typically been applied to either
elliptic, parabolic, or mixed type problems. The key ingredient 
of a stable dG scheme is an appropriate choice of numerical flux, 
and our particular choice has been motivated by the analysis
presented in Appendix \ref{WPToy}. When used to evolve the
Schwarzschild solution in Kerr-Schild coordinates, our numerical
implementation of the BSSN system \eqref{eqn:GBSSNL_spherical} 
is robustly stable and converges to the analytic solution 
exponentially with increased polynomial order. By approximating the 
spatially second-order form of the BSSN system, we have not introduced
extra fields which are {\em evolved}. Evolved auxiliary fields
result in new constraints which may spoil stability. 
Our main goal has been stable evolution of the spherically 
reduced BSSN system
as a first step towards understanding how a discontinuous
Galerkin method might be applied to the full BSSN system.
Towards that goal, we now discuss treatment of singularities 
and generalization of the described dG method to higher space 
dimension.

To deal with the fixed Schwarzschild singularity, we have used 
excision which is easy in the context of the spherically reduced 
BSSN system. However, excision for the binary black hole problem 
in full general relativity requires attention to the technical 
challenge of horizon tracking. State-of-the-art BSSN codes avoid 
such complication, relying instead on the moving-puncture technique.
While the moving-puncture technique does involve 
mild central singularities, it may still 
prove amenable to spectral methods. Indeed, spectral methods for 
non-smooth problems is 
well-developed in both theory and for complex applications. 
Since the moving-puncture technique can be performed in spherical 
symmetry \cite{BrownSphSymBSSN}, a first-step toward a spectral 
moving-puncture code would be to implement a moving puncture 
with the nodal dG method described here. Such an 
implementation may adopt Legendre-Gauss-Radau nodes on 
the innermost subdomain, thereby ensuring that the physical 
singularity does not lie on a nodal point (in much the same way 
finite difference codes use a staggered grid). 
Beyond traditional excision and moving punctures,
one might construct smooth initial data via the turducken approach 
to singularities. However, in combination with 1+$\log$ slicing 
and the Gamma-driver shift condition, turduckened initial data will 
evolve towards a ``trumpet" geometry 
\cite{Brown_Turducken1,Brown_Turducken2}.
%Thus one may be generically confronted with mildly singular 
%behavior near a physical singularity
%regardless of the initial data's smoothness.

Discontinuous Galerkin methods for hyperbolic problems in two and
three space dimensions are well-developed. A generalization of the 
method described here to three-dimensions and the full BSSN system
would likely rely on an unstructured mesh.
Appropriate local polynomial expansions for the subelements 
are well-understood, as are choices for the numerical
fluxes which would now live on two-dimensional faces 
rather than single points. Whether or not it would ultimately 
prove successful, generalization of our dG method to a higher 
dimension would rely on an established conceptual framework.
Further computational advances of relevance to a generalization 
of our dG method to the full BSSN system (possibly including matter)
may include mesh $hp$-adaptivity, local timestepping, shock 
capturing and slope limiting techniques \cite{HEST}. Moreover, 
recent work \cite{KlocHestWar} indicates that enhanced performance 
would be expected were our scheme implemented
on graphics processor units.

\section{Acknowledgments}\label{sec:acknowledgments}

We thank Nick Taylor for discussions about treating second-order 
operators with spectral methods, Benjamin Stamm for helping 
polish up a few parts of Appendix \ref{WPToy}, Khosro Shahbazi 
for discussions on LDG and IP methods. We also acknowledge 
helpful conversations with David Brown and Manuel Tiglio about 
the BSSN system and previous work on its numerical implementation.
We gratefully acknowledge funding through grants DMS 0554377 
and DARPA/AFOSR FA9550-05-1-0108 to Brown University and
NSF grant PHY 0855678 to the University of New Mexico.

\appendix
%%%%%%%%%%%%%%%%%%%%%%%%%%%%%%%%%%%%%%%%%%%%%%%%%%%%%%%%%%%%%%%
%
%  APPENDIX: HYPERBOLICITY OF first-order SYSTEM.
%
%%%%%%%%%%%%%%%%%%%%%%%%%%%%%%%%%%%%%%%%%%%%%%%%%%%%%%%%%%%%%%%
\section{Hyperbolicity of the first-order 
system.}\label{app:hyperbolic}
This appendix analyzes the matrix $\mathcal{A}(u)$ appearing in
\eqref{eqn:firstordersys_abs} in order to construct the 
characteristic fields 
\eqref{eqn:characteristic_fields}. In matrix form the sector 
\eqref{eqn:GBSSNPrinciplePart} of the principal part of 
\eqref{eqn:firstordersys_abs} reads as follows:
\begin{equation}\label{eqn:matrix_pp}
\partial_t
\left[
\begin{array}{c}
B^r\\
\\
A_{rr}\\
\\
K\\
\\
\Gamma^r\\
\\
Q_\chi\\
\\
Q_{g_{rr}}\\
\\
Q_{g_{\theta\theta}}\\
\\
Q_{\alpha}\\
\\
Q_{\beta^r}\\
\end{array}
\right]
=
\left[
\begin{array}{ccccccccc}
%
%  ROW1
%
\beta^r                                         & 0                              & -\frac{4\lambda\alpha}{3g_{rr}}      & 
0                                               & 0                              &  \frac{\lambda\beta^r}{6(g_{rr})^2}  & 
\frac{\lambda\beta^r}{3g_{\theta\theta} g_{rr}} & 0                              &  \frac{4\lambda}{3g_{rr}}            \\
%%%%%%%%%%%%%%%%%%%%%%%%%%%%%%%%%%%%%%%%%%%%%%%%%%%%%%%%%%%%%%%%%%%%%%%%%%%%%%%%%%%%%%%%%%%%%%%%%%%%%%%%%%%%%%%%%%%%%%%%%%
&              &              &              &              &              &              &              &              \\
%%%%%%%%%%%%%%%%%%%%%%%%%%%%%%%%%%%%%%%%%%%%%%%%%%%%%%%%%%%%%%%%%%%%%%%%%%%%%%%%%%%%%%%%%%%%%%%%%%%%%%%%%%%%%%%%%%%%%%%%%%
%
%  ROW2
%

0                                               & \beta^r                        & 0                                    &
\frac{2}{3}g_{rr}\alpha \chi                    & \frac{1}{3}\alpha              & -\frac{\alpha\chi}{3g_{rr}}          &
 \frac{\alpha\chi}{3g_{\theta\theta}}           & -\frac{2}{3}\chi               & 0                                    \\
%%%%%%%%%%%%%%%%%%%%%%%%%%%%%%%%%%%%%%%%%%%%%%%%%%%%%%%%%%%%%%%%%%%%%%%%%%%%%%%%%%%%%%%%%%%%%%%%%%%%%%%%%%%%%%%%%%%%%%%%%%
&              &              &              &              &              &              &              &              \\
%%%%%%%%%%%%%%%%%%%%%%%%%%%%%%%%%%%%%%%%%%%%%%%%%%%%%%%%%%%%%%%%%%%%%%%%%%%%%%%%%%%%%%%%%%%%%%%%%%%%%%%%%%%%%%%%%%%%%%%%%%
%
%  ROW3
%
0                                               & 0                              & \beta^r                              &
0                                               & 0                              & 0                                    &
0                                               & -\frac{\chi}{g_{rr}}           & 0                                    \\
%%%%%%%%%%%%%%%%%%%%%%%%%%%%%%%%%%%%%%%%%%%%%%%%%%%%%%%%%%%%%%%%%%%%%%%%%%%%%%%%%%%%%%%%%%%%%%%%%%%%%%%%%%%%%%%%%%%%%%%%%%
&              &              &              &              &              &              &              &              \\
%%%%%%%%%%%%%%%%%%%%%%%%%%%%%%%%%%%%%%%%%%%%%%%%%%%%%%%%%%%%%%%%%%%%%%%%%%%%%%%%%%%%%%%%%%%%%%%%%%%%%%%%%%%%%%%%%%%%%%%%%%
%
%  ROW 4
%
0                                               & 0                              & -\frac{4\alpha}{3g_{rr}}             &
\beta^r                                         & 0                              & \frac{\beta^r}{6(g_{rr})^2}           & 
\frac{\beta^r}{3 g_{\theta\theta} g_{rr}}       & 0                              & \frac{4}{3g_{rr}}                    \\
%%%%%%%%%%%%%%%%%%%%%%%%%%%%%%%%%%%%%%%%%%%%%%%%%%%%%%%%%%%%%%%%%%%%%%%%%%%%%%%%%%%%%%%%%%%%%%%%%%%%%%%%%%%%%%%%%%%%%%%%%%
&              &              &              &              &              &              &              &              \\
%%%%%%%%%%%%%%%%%%%%%%%%%%%%%%%%%%%%%%%%%%%%%%%%%%%%%%%%%%%%%%%%%%%%%%%%%%%%%%%%%%%%%%%%%%%%%%%%%%%%%%%%%%%%%%%%%%%%%%%%%%
%
%  ROW 5
%
0                                               & 0                              & \frac{2}{3}\alpha\chi                &
0                                               & \beta^r                        & -\frac{\beta^r\chi}{3g_{rr}}         &
-\frac{2\beta^r\chi}{3g_{\theta\theta}}         & 0                              & -\frac{2}{3}\chi                     \\
%%%%%%%%%%%%%%%%%%%%%%%%%%%%%%%%%%%%%%%%%%%%%%%%%%%%%%%%%%%%%%%%%%%%%%%%%%%%%%%%%%%%%%%%%%%%%%%%%%%%%%%%%%%%%%%%%%%%%%%%%%
&              &              &              &              &              &              &              &              \\
%%%%%%%%%%%%%%%%%%%%%%%%%%%%%%%%%%%%%%%%%%%%%%%%%%%%%%%%%%%%%%%%%%%%%%%%%%%%%%%%%%%%%%%%%%%%%%%%%%%%%%%%%%%%%%%%%%%%%%%%%%
%
%  ROW 6
%
0                                               & -2\alpha                       & 0                                    &
0                                               &  0                             & \frac{2}{3}\beta^r                   &
-\frac{2g_{rr}\beta^r}{3g_{\theta\theta}}       &  0                             & \frac{4}{3}g_{rr}                    \\
%%%%%%%%%%%%%%%%%%%%%%%%%%%%%%%%%%%%%%%%%%%%%%%%%%%%%%%%%%%%%%%%%%%%%%%%%%%%%%%%%%%%%%%%%%%%%%%%%%%%%%%%%%%%%%%%%%%%%%%%%%
&              &              &              &              &              &              &              &              \\
%%%%%%%%%%%%%%%%%%%%%%%%%%%%%%%%%%%%%%%%%%%%%%%%%%%%%%%%%%%%%%%%%%%%%%%%%%%%%%%%%%%%%%%%%%%%%%%%%%%%%%%%%%%%%%%%%%%%%%%%%%
%
%  ROW 7
%
0                                    & \frac{g_{\theta\theta}\alpha}{g_{rr}} & 0                                        &
0                                    & 0                                     & -\frac{g_{\theta\theta}\beta^r}{3g_{rr}} &
\frac{1}{3}\beta^r                   & 0                                     & -\frac{2}{3}g_{\theta\theta}             \\
%%%%%%%%%%%%%%%%%%%%%%%%%%%%%%%%%%%%%%%%%%%%%%%%%%%%%%%%%%%%%%%%%%%%%%%%%%%%%%%%%%%%%%%%%%%%%%%%%%%%%%%%%%%%%%%%%%%%%%%%%%
&              &              &              &              &              &              &              &              \\
%%%%%%%%%%%%%%%%%%%%%%%%%%%%%%%%%%%%%%%%%%%%%%%%%%%%%%%%%%%%%%%%%%%%%%%%%%%%%%%%%%%%%%%%%%%%%%%%%%%%%%%%%%%%%%%%%%%%%%%%%%
%
%  ROW 8
%
0                                               & 0                              & -2\alpha                             &
0                                               & 0                              &  0                                   &
0                                               & \beta^r                        &  0                                   \\
%%%%%%%%%%%%%%%%%%%%%%%%%%%%%%%%%%%%%%%%%%%%%%%%%%%%%%%%%%%%%%%%%%%%%%%%%%%%%%%%%%%%%%%%%%%%%%%%%%%%%%%%%%%%%%%%%%%%%%%%%%
&              &              &              &              &              &              &              &              \\
%%%%%%%%%%%%%%%%%%%%%%%%%%%%%%%%%%%%%%%%%%%%%%%%%%%%%%%%%%%%%%%%%%%%%%%%%%%%%%%%%%%%%%%%%%%%%%%%%%%%%%%%%%%%%%%%%%%%%%%%%%
%
%  ROW 9
%
\frac{3}{4}                                     & 0                              & 0                                    &
0                                               & 0                              & 0                                    &
0                                               & 0                              & \beta^r                              
\end{array}
\right]
\left[
\begin{array}{c}
B^r\\
\\
A_{rr}\\
\\
K\\
\\
\Gamma^r\\
\\
Q_\chi\\
\\
Q_{g_{rr}}\\
\\
Q_{g_{\theta\theta}}\\
\\
Q_\alpha\\
\\
Q_{\beta^r} 
\end{array}
\right]^{{}_{\mbox{\LARGE $'$}}},
\end{equation}
which defines the matrix $\tilde{A}(u)$ appearing 
in \eqref{eqn:partofpp}, and so also the matrix
$\mathcal{A}(u)$ in \eqref{eqn:firstordersys_abs}.
Note that in the last equation the matrix within the 
square brackets is $-\tilde{A}(u)$.
For certain configurations of $u$ and $\lambda$, the 
system \eqref{eqn:firstordersys_abs} is strongly hyperbolic 
\cite{KreissLorenz}, that is $\mathcal{A}(u)$ has a complete 
set of eigenvectors and real eigenvalues. Indeed, five eigenpairs 
of $\mathcal{A}(u)$ are trivially recovered upon inspection of 
$\mathcal{A}(u)$'s leading $5\times 5$ diagonal block. These
correspond to eigenvalue 0 and the left eigenspace 
$\{\xi_j = e_j^T : 1 \leq j \leq 5\}$, where $e_j$ are the 
canonical basis vectors. Since each component of $u$ arises as
$e_j^T W$, each is also a characteristic field.

The remaining nine eigenpairs are determined by $\tilde{A}(u)$. 
The eigenvalues of $\tilde{A}(u)$ are
\begin{equation}\label{eqn:mu_evalues}
\mu_1 = 0,
\quad
\mu_{2,3} = -\beta^r,
\quad
\mu_4^\pm = -\beta^r \pm \sqrt{\frac{2\alpha\chi}{g_{rr}}},
\quad
\mu_5^\pm = -\beta^r \pm \alpha \sqrt{\frac{\chi}{g_{rr}}},
\quad
\mu_6^\pm = -\beta^r \pm \sqrt{\frac{\lambda}{g_{rr}}},
\end{equation}
and the corresponding left eigenvectors are
\begin{subequations}
\begin{align}
x_{1}     & = (0,0,0,0,0,g_{\theta\theta},2g_{rr},0,0) \\
x_{2}     & = \left(0,0,0,g_{rr},
                  \frac{2}{\chi},-\frac{1}{2g_{rr}},
                 -\frac{1}{g_{\theta\theta}},0,0\right) \\
x_{3}     & = \left(\frac{g_{rr}}{\lambda},0,0,0,
                  \frac{2}{\chi},-\frac{1}{2g_{rr}},
                 -\frac{1}{g_{\theta\theta}},0,0\right) \\
x_{4}^\pm & = \left(0,0,\pm\sqrt{\frac{2\alpha g_{rr}}{\chi}},
                   0,0,0,0,1,0\right)\\
x_{5}^\pm & = \left(0,\mp\frac{3}{\sqrt{g_{rr}\chi}},
                  \pm 2\sqrt{\frac{g_{rr}}{\chi}},
                  2g_{rr},
                  \frac{1}{\chi},
                 -\frac{1}{g_{rr}},
                  \frac{1}{g_{\theta\theta}},
                  0,0\right) \\
x_{6}^\pm & = \left(-\frac{3}{4}\frac{g_{rr}}{\lambda},
                  0,\pm\frac{\alpha\sqrt{\lambda g_{rr}}}{(2\alpha\chi-\lambda)},
                  0,0,-\frac{\beta^r}{8(\beta^r g_{rr}\mp
                  \sqrt{\lambda g_{rr}})},
                  \right. \nonumber 
\\ 
          & \hspace{5mm}
                 \left. -\frac{\beta^rg_{rr}}{4g_{\theta\theta}
                  (\beta^r g_{rr}
                  \mp\sqrt{\lambda g_{rr}})},
                  \frac{\alpha\chi}{(2\alpha\chi-\lambda)},
                  \pm\sqrt{\frac{g_{rr}}{\lambda}}
\right), 
\end{align}
\end{subequations}
where for example $x_{5}^\pm \tilde{A}(u) = \mu_5^\pm x_{5}^\pm$.
Assuming that $g_{rr}$, $g_{\theta\theta}$,
$\chi$, and $\alpha$ are everywhere strictly positive,
the eigenvalues are real and the eigenvectors are linearly 
independent provided that \eqref{eqn:hyperbolicity_cond} holds.
These eigenvectors are easily extended to eigenvectors of 
$\mathcal{A}(u)$, e.~g.~as
$x_{6}^\pm \rightarrow (0_{1\times 5},x_6^\pm)$. Then, for example, 
the characteristic field
\begin{equation}
X_6^\pm \equiv  (0_{1\times 5},x_6^\pm) W = x_6^\pm W_{v:Q},
\end{equation}
and similarly $X_j^\pm = x^\pm_j W_{v:Q}$ for $j=4,5$ and
$X_k = x_k W_{v:Q}$ for $k = 1,2,3$. 
The characteristic speeds for these fields are $\mu_k$ 
and $\mu^\pm_j$. With this convention the speeds 
listed in Table \ref{tab:speeds} correspond to 
the $X_k$ and $X_j^\pm$ in 
\eqref{eqn:characteristic_fields}.

%%%%%%%%%%%%%%%%%%%%%%%%%%%%%%%%%%%%%%%%%%%%%%%%%%%%%%%%%%%%%%%
%
%  APPENDIX: KERR-SCHILD
%
%%%%%%%%%%%%%%%%%%%%%%%%%%%%%%%%%%%%%%%%%%%%%%%%%%%%%%%%%%%%%%%
\section{Schwarzschild solution in conformal 
Kerr-Schild coordinates.} \label{app:KerrSchild}
In Kerr-Schild coordinates, here the system directly related 
to {\em incoming} Eddington-Finkelstein null coordinates, 
the line element for the Schwarzschild solution reads
\begin{equation}
ds^2 = -\alpha^2 dt^2 + (1+ 2M/R) (dR + \beta^R dt)^2
+ R^2 d\theta^2
+ R^2 \sin^2\theta d\phi^2,
\end{equation}
where $R$ is the area radius, $\alpha = (1+2M/R)^{-1/2}$
is the lapse, and $\beta^R = 2M/(R+2M)$ is the shift vector.
The physical spatial metric $\bar{g}_{ab}$ is the spatial part 
of this line element.

To define the corresponding solution to the BSSN system, 
we use equation $g_{ab} = \chi \bar{g}_{ab}$ to define
the following relationship between line elements:
\begin{equation}
dr^2 + r^2 (d\theta^2 + \sin^2\theta d\phi^2)
= \chi [(1+ 2M/R) dR^2 
+ R^2 d\theta^2 
+ R^2 \sin^2\theta d\phi^2 ],
\end{equation}
so that
\begin{equation}\label{eq:comparelineelements}
\chi\left(1 + \frac{2M}{R}\right)
\left(\frac{dR}{dr}\right)^2 = 1,
\qquad
\chi R^2 = r^2.
\end{equation}
Then we have
\begin{equation}
\left(1+\frac{2M}{R}\right)^{1/2}\frac{dR}{R}
= \frac{dr}{r},
\end{equation}
with integration yielding
\begin{equation}
r = \frac{R}{4}\left(1 
+ \sqrt{1 + \frac{2M}{R}}\right)^2
e^{2-2\sqrt{1 + 2M/R}},
\end{equation}
where the constant of integration has been chosen
so that the $R,r \rightarrow \infty$ limits are
consistent. The second relation in 
\eqref{eq:comparelineelements} shows that
\begin{equation}
\chi = \frac{1}{16}\left(1
+ \sqrt{1 + \frac{2M}{R}}\right)^4
e^{4-4\sqrt{1 + 2M/R}},
\qquad	
\chi^{-4} = \frac{
2e^{\sqrt{1 + 2M/R}-1}}{1 
+ \sqrt{1 + 2M/R}}.
\end{equation}
The extrinsic curvature tensor is specified by the
expression for $K$ given in (\ref{eq:BSSN_KS}h),
the identity $K = K^R_R + 2K^\theta_\theta$, and
\begin{align}
K^\theta_\theta = 
\left(1+\frac{2M}{R}\right)^{-1/2}
    \frac{2M}{R^2}.
\end{align}
Since $K^R_R = K^r_r$, we compute that
\begin{align}
K^r_r = K - 2 K^\theta_\theta
= -\left(1+\frac{2M}{R}\right)^{-1/2}
  \left(
  \frac{R+M}{R+2M}
  \right)
  \frac{2M}{R^2}.
\end{align}
Next, since $K_{rr} = \bar{g}_{rr} K^r_r = \chi^{-1} K^r_r$, we
have $K^r_r = A_{rr} + {\textstyle \frac{1}{3}} g_{rr} K$. This
implies $A_{rr} = K^r_r - {\textstyle \frac{1}{3}} K$, from
which we get (\ref{eq:BSSN_KS}g). In all we have
\begin{subequations}\label{eq:BSSN_KS}
\begin{align}
\alpha & = \left(1+\frac{2M}{R}\right)^{-1/2}\\
\beta^r & = \beta^R \frac{dr}{dR} = \chi^{1/2}
\left(1+\frac{2M}{R}\right)^{-1/2}\frac{2M}{R}\\ 
g_{rr} & = 1\\
g_{\theta\theta} & = r^2 = \chi R^2\\
\chi & = \frac{1}{16}\left(1
+ \sqrt{1 + \frac{2M}{R}}\right)^4
e^{4-4\sqrt{1 + 2M/R}}\\
B^r & = 0 \\
A_{rr} & =
-\left(1+\frac{2M}{R}\right)^{-1/2}\frac{4M}{3R^2}
\left(\frac{2R+3M}{R+2M}
\right)
\\
K & = \left(1+\frac{2M}{R}\right)^{-3/2}
\left(1+\frac{3M}{R}\right) \frac{2M}{R^2}\\
\Gamma^r & = -\frac{2}{r} = -\frac{2}{\chi^{1/2}R}.
\end{align}
\end{subequations}
To differentiate these expressions with respect to $r$, we use
the identity
\begin{equation}
\frac{dR}{dr} = \chi^{-1/2}\left(1 + \frac{2M}{R}\right)^{-1/2}
\end{equation}
along with the chain rule.

%%%%%%%%%%%%%%%%%%%%%%%%%%%%%%%%%%%%%%%%%%%%%%%%%%%%%%%%%%%%%%%
%
%  APPENDIX: STABILITY OF TOY PDE
%
%%%%%%%%%%%%%%%%%%%%%%%%%%%%%%%%%%%%%%%%%%%%%%%%%%%%%%%%%%%%%%%
\section{Stability of the model system}\label{WPToy}
The following stability analysis for the model
system \eqref{eqn:TOY} has been inspired by
\cite{ShuSchro,ShuKvd}. After dropping all nonlinear source 
terms, the system 
\eqref{eqn:TOY} becomes
\begin{subequations} \label{eqn:TOY_Lin}
\begin{align}
\partial_t u & =  u' + av \\ 
\partial_t v & =  u'' + v'.
\end{align}
\end{subequations}
This section analyzes the stability of \eqref{eqn:TOY_Lin}, 
considering both the continuum system itself as well as its 
semi-discrete dG approximation. The latter analysis offers 
some insight into the empirically observed stability of our 
dG scheme for the spherically reduced BSSN equations.

\subsection{Analysis for a single interval}
Throughout we work with the $L_2$ inner product and norm,
\begin{equation}
(f,g)_\domain = \int_{\domain} fg,\qquad \|f\|_{\domain} 
= \sqrt{(f,f)_\domain},
\end{equation}
where $\domain$ is the spatial coordinate interval 
(here $\domain$ may represent a subdomain
$\domain^k$ or the whole domain $\Omega$), and 
we have suppressed all integration measures. For the 
continuum model we will establish the following estimate:
\begin{align} \label{eqn:L2Bound}
\|u'(T,\cdot)\|^2_{\domain} + a\|v(T,\cdot)\|^2_{\domain}
\leq C(T)
\left(\|u'(0,\cdot)\|^2_{\domain} 
+ a\|v(0,\cdot)\|^2_{\domain} \right),
\end{align}
where the time-dependent constant $C(T)$ is determined
solely by the choice of boundary conditions. 
To show \eqref{eqn:L2Bound}, we first change 
variables with $\hat{v} = \sqrt{a}v$,  
thereby rewriting \eqref{eqn:TOY_Lin} in
the following symmetric form:
\begin{subequations} \label{eqn:TOY_Lin_Mod}
\begin{align}
\partial_t u & =  u' + \sqrt{a}\hat{v} \\ 
\partial_t \hat{v} & =  \sqrt{a}u'' + \hat{v}'.
\end{align}
\end{subequations}
Equations (\ref{eqn:TOY_Lin_Mod}a,b) then imply
\begin{subequations}\label{uprimeANDv_energy}
\begin{align}
\frac{1}{2}
\partial_t \int_{\domain} (u')^2 
& = \int_{\domain} u'(u'' + \sqrt{a}\hat{v}') 
  = \int_{\domain} \sqrt{a} u' \hat{v}' 
  + \frac{1}{2} \int_{\partial \domain} (u')^2
\\
\frac{1}{2}
    \partial_t \int_{\domain}(\hat{v})^2 
& = \int_{\domain}\hat{v} (\sqrt{a}u''+\hat{v}') 
  = -\int_{\domain}\sqrt{a}u'\hat{v}'
  + \frac{1}{2}
    \int_{\partial \domain}(
    \hat{v}{}^2
  + 2\sqrt{a} u' \hat{v}).
\end{align}
\end{subequations}
Here $\hat{v}\hat{v}'$ and $u'u''$ have been expressed as
exact derivatives and then integrated to boundary terms,
the second equation employs an extra integration
by parts, and with only one space dimension 
$\int_{\partial \domain}$ denotes a difference of 
endpoint evaluations. Addition of 
Eqs.~(\ref{uprimeANDv_energy}a,b) gives
\begin{align}
\frac{1}{2}
\partial_t \int_{\domain} 
\big[\hat{v}{}^2 + (u')^2\big] 
=
\frac{1}{2} 
\int_{\partial \domain}
\big[\hat{v}{}^2 + (u')^2 + 2\sqrt{a} u'\hat{v} \big].
\end{align}
Substitutions with the identities
\begin{align}
\big[\hat{v}{}^2 + (u')^2\big] = 
\frac{1}{2}
\big[(\hat{v} + u')^2 + (\hat{v} - u')^2\big],\qquad
2 u'\hat{v} = 
\frac{1}{2}
\big[(\hat{v} + u')^2 - (\hat{v} - u')^2\big]
\end{align}
and replacements to recover the original 
variable $v = \hat{v}/\sqrt{a}$ yield
\begin{align}\label{eq:d_tenergy}
\frac{1}{2}
\partial_t \int_{\domain}
\big[a v^2 + (u')^2\big]
= 
\frac{1+\sqrt{a}}{4}
\int_{\partial \domain} 
( \sqrt{a} v + u' )^2 + 
\frac{1-\sqrt{a}}{4}
\int_{\partial \domain} 
( \sqrt{a} v - u')^2.
\end{align}
From \eqref{eq:d_tenergy} we deduce that the time-dependent 
constant $C(T)$ in \eqref{eqn:L2Bound} must satisfy
\begin{equation}
\left|1 + \frac{\int_0^T\left[
\frac{1+\sqrt{a}}{2}
\int_{\partial \domain}
( \sqrt{a} v + u' )^2 +
\frac{1-\sqrt{a}}{2}
\int_{\partial \domain}
( \sqrt{a} v - u')^2
\right]dt}{
\|u'(0,\cdot)\|^2 + a \|v(0,\cdot)\|^2} \right|
\leq C(T).
\end{equation}
For periodic boundary conditions, we may choose $C(T) = 1$.
Moreover, if $a \geq 1$ and $u' = -\sqrt{a} v$ is specified at 
$\partial\domain^+$, then 
$\|u'(t,\cdot)\|^2 + a \|v(t,\cdot)\|^2$ decays.

Still working on a single interval (subdomain), we now 
consider the semi-discrete scheme for 
\eqref{eqn:TOY_Lin_Mod},
i.~e.~\eqref{eqn:TOY_SD} with all nonlinear source 
terms dropped, and with $v$ replaced by $\hat{v}/\sqrt{a}$. 
Derivation of a formula analogous to \eqref{eq:d_tenergy} is
our first step toward establishing $L_2$ stability of 
the semi-discrete scheme. While \eqref{eqn:TOY_SD} features
vectors, for example $\boldsymbol{u}(t)$, taking values at 
the Legendre-Gauss-Lobatto nodal points, here we work 
with the numerical solution as a polynomial, for example 
$u_h(t,x)$. These two representations are related
by the Lagrange interpolating polynomials for the nodal
set, here taken to span both the space of test functions 
and the space of basis functions. Our scheme is
\begin{subequations} \label{eqn:Toy_SD_WP}
\begin{align}
\int_{\mathsf{D}^k} \psi \partial_t u_h & = 
\int_{\mathsf{D}^k} \psi (Q_h + \sqrt{a}\hat{v}_h) 
\\ 
\int_{\mathsf{D}^k} \xi \partial_t \hat{v}_h & =
-\int_{\mathsf{D}^k} \xi'(\sqrt{a}Q_h + \hat{v}_h) 
+\int_{\partial \mathsf{D}^k} \xi (\sqrt{a}Q^* + \hat{v}{}^*)\\
\int_{\mathsf{D}^k} \varphi Q_h  & =
\int_{\mathsf{D}^k} \varphi u_h' 
+ \int_{\partial \mathsf{D}^k} \varphi \left(u^* - u_h\right),
\end{align}
\end{subequations}
where $\psi$, $\xi$, and $\varphi$ are polynomial test functions. 
These test functions are arbitrary, except that they 
must be degree-$N$ polynomials. In \eqref{eqn:Toy_SD_WP} the
variables $u_h$, $\hat{v}_h$ and $Q_h$ should also carry a
superscript $k$, but we have suppressed this.
Derivation of a formula analogous to \eqref{eq:d_tenergy} is 
complicated by the fact that $Q_h$ is not evolved.
Nevertheless, at a given instant $t$ we can assemble $Q_h$ 
from (\ref{eqn:Toy_SD_WP}c). 

Mimicking the calculation (\ref{uprimeANDv_energy}b) from 
the continuum case, we first use (\ref{eqn:Toy_SD_WP}b) with 
$\xi=\hat{v}_h$ to write
\begin{align} \label{eqn:V_Energy}
\begin{split}
\frac{1}{2} \partial_t \int_{\mathsf{D}^k} \hat{v}_h^2 
& = 
-\int_{\mathsf{D}^k}
(\sqrt{a}Q_h + \hat{v}_h)\hat{v}_h' 
+\int_{\partial \mathsf{D}^k}
(\sqrt{a}Q^* + \hat{v}{}^*)\hat{v}_h  \\
& = -\int_{\mathsf{D}^k} \sqrt{a}Q_h \hat{v}_h'
+ \frac{1}{2}
\int_{\partial \mathsf{D}^k} 
\big[2(\sqrt{a}Q^* + \hat{v}{}^*)\hat{v}_h 
- \hat{v}_h^2\big].
\end{split}
\end{align}
The right-hand side of (\ref{uprimeANDv_energy}a) is 
analogous to
\begin{align}
\frac{1}{2} \partial_t \int_{\mathsf{D}^k} Q_h^2 
= \int_{\mathsf{D}^k} Q_h \partial_t Q_h .
\end{align}
However, since $Q_h$ is not evolved, the term $\partial_t Q_h$
must be given a suitable interpretation. 
On the right hand side of (\ref{eqn:Toy_SD_WP}c) only 
$u_h$, $u_h'$, and $u^*$ necessarily depend on time, since 
the test function $\varphi$ need not be time-dependent. 
Furthermore, $u^*$ is explicitly given as a linear combination 
of $u_h$, as seen in Eq.~\eqref{eqn_nf_u} below. 
Choosing $\varphi = \ell_j$,
taking the time derivative of (\ref{eqn:Toy_SD_WP}c), and 
appealing to the commutivity of mixed partial derivatives, 
we therefore arrive at
\begin{align}
\int_{\mathsf{D}^k}  \ell_j \partial_t Q_h 
= \int_{\mathsf{D}^k} \ell_j (\partial_t u_h)' 
+ \int_{\partial \mathsf{D}^k} \ell_j
\big((\partial_t u)^* - \partial_t u_h\big),
\end{align}
where $(\partial_t u)^*$ depends on $\partial_t u_h$
in precisely the same way that $u^*$ depends on $u_h$. We have 
written $\ell_j$ rather than $\varphi$ in the last equation
to emphasize that the result also holds for any linear 
combination of $\ell_j$ (for example $\varphi$), and even
for {\em time-dependent} combinations. Since $Q_h$ is itself
such a combination, we obtain
\begin{align}
\begin{split}\label{eqn_Q_energy}
\frac{1}{2} \partial_t \int_{\mathsf{D}^k} Q_h^2 
& = \int_{\mathsf{D}^k}  Q_h (\partial_t u_h)' 
+ \int_{\partial \mathsf{D}^k} 
\big((\partial_t u)^* - \partial_t u_h\big) Q_h 
\\
& = \int_{\mathsf{D}^k}  Q_h  
(Q_h' + \sqrt{a}\hat{v}_h') 
+\int_{\partial \mathsf{D}^k} 
\big(
(\partial_t u)^* - \partial_t u_h\big)
Q_h 
\\
& = 
\int_{\mathsf{D}^k}  \sqrt{a} Q_h \hat{v}_h'
+ \frac{1}{2}\int_{\partial \mathsf{D}^k} 
\big[2((\partial_t u)^* - \partial_t u_h)Q_h 
+ Q_h^2
\big].
\end{split}
\end{align}
Addition of \eqref{eqn:V_Energy} and \eqref{eqn_Q_energy}
gives
\begin{align}\label{eq:semides_result}
\frac{1}{2} \partial_t \int_{\mathsf{D}^k} 
(Q_h^2 + \hat{v}_h^2) 
= \frac{1}{2}
\int_{\partial \mathsf{D}^k} 
\big[
Q_h^2 - \hat{v}_h^2
+ 2(\sqrt{a}Q^* + \hat{v}{}^*)\hat{v}_h + 
  2((\partial_t u)^* - \partial_t u_h)Q_h\big],
\end{align}
the aforementioned analog of \eqref{eq:d_tenergy}. This
formula holds on a single subdomain $\mathsf{D}^k$, and
we now combine multiple copies of it, one for each value of $k$.

\subsection{Analysis for multiple intervals}
To facilitate combination of \eqref{eq:semides_result} over
all $k$, we change notation. At every subdomain interface 
$\mathsf{I}^{k+1/2} 
\equiv \partial \mathsf{D}^k \cap \partial \mathsf{D}^{k+1}$, 
let the superscripts $L$ and $R$ denote field values 
respectively taken from the left and right. 
Then the fields 
evaluated at $\mathsf{I}^{k+1/2}$ which belong to $\mathsf{D}^k$ 
will be $u_{k+1/2}^L$, $\hat{v}_{k+1/2}^L$, and 
$Q_{k+1/2}^L$, while those belonging to $\mathsf{D}^{k+1}$ will 
be $u_{k+1/2}^R$, $\hat{v}_{k+1/2}^R$, and	
$Q_{k+1/2}^R$. However, at $\mathsf{I}^{k-1/2}$
the values taken from $\mathsf{D}^k$ are 
$u_{k-1/2}^R$, $\hat{v}_{k-1/2}^R$, and
$Q_{k-1/2}^R$. Note that we have also replaced
the subscript $h$, denoting a numerical solution, with 
$k\pm1/2$, denoting the location of the endpoint value
of the numerical solution. With this notation, we define
\begin{align}\label{eqn:DeltaLalpha}
\Delta^L_\alpha
= \frac{1}{2} \left[ (Q^L_\alpha)^2 -
(\hat{v}^L_\alpha)^2\right]
+ \left(\sqrt{a}Q_\alpha^* + 
\hat{v}_\alpha^*\right)\hat{v}^L_\alpha
+ \left[(\partial_t u_\alpha)^* - \partial_t 
u_\alpha^L\right]Q^L_\alpha,
\end{align}
and similarly for $\Delta^R_\alpha$. The same 
numerical fluxes appear in both $\Delta^{L}_\alpha$ and
$\Delta^{R}_\alpha$ (i.e.~each numerical flux takes
the same value on either side of an interface), 
whence fluxes like $Q^*_\alpha$ do not carry an $L$ or 
$R$ superscript. In terms of these definitions 
\eqref{eq:semides_result} becomes
\begin{align}
\frac{1}{2} \partial_t \int_{\mathsf{D}^k}
(Q_h^2 + \hat{v}_h^2) = \Delta^{L}_{k+1/2} - \Delta^R_{k-1/2}.
\end{align}
Summation over all $\mathsf{D}^k$ yields
\begin{align}\label{eq:ksummation}
\frac{1}{2} \partial_t \sum_{k=1}^{k_\mathrm{max}}
\int_{\mathsf{D}^k}
(Q_h^2 + \hat{v}_h^2) & = 
\sum_{k=1}^{k_\mathrm{max}-1} 
\big(\Delta^{L}_{k+1/2} - \Delta^R_{k+1/2} \big) 
+ \Delta^{L}_{k_\mathrm{max}+1/2} - \Delta^R_{1/2}
\nonumber \\
& =
\sum_{k=1}^{k_\mathrm{max}-1}
\big(\Delta^{L}_h - \Delta^R_h
\big)\big|_{\mathsf{I}^{k+1/2}}
+ \Delta^{L}_{k_\mathrm{max}+1/2} - \Delta^R_{1/2}.
\end{align}
We have reverted to $h$-notation denoting the
numerical solution, since the $L,R$ superscripts indicate unambiguously
the relevant domain used for evaluation at $\mathsf{I}^{k+1/2}$. 

We again seek an estimate of the form
\begin{align} \label{eqn:NumBound}
    \sum_{k=1}^{k_\mathrm{max}}\big(\|Q_h(T,\cdot)\|^2_{\domain^k}
+ a \|v_h(T,\cdot)\|^2_{\domain^k} \big)
    \leq C(T) \sum_{k=1}^{k_\mathrm{max}} \big(
    \|Q_h(0,\cdot)\|_{\domain^k}^2
+ a \|v_h(0,\cdot)\|_{\domain^k}^2 \big),
\end{align}
that is essentially the same as the one
\eqref{eqn:L2Bound} considered in the continuum case.
Assume that the chosen boundary conditions 
ensure $\Delta^{L}_{k_\mathrm{max}+1/2} - \Delta^R_{1/2}$ 
is bounded by a time-dependent constant which does not 
depend on the numerical parameters $N$ and $h$ 
(subdomain width). 
Establishment of stability then amounts to showing 
that the remaining sum over interface terms in 
\eqref{eq:ksummation} is non-positive; whence 
this remaining sum is consistent with $C(T) \leq 1$,
although the boundary conditions may give rise to 
a different bound.
In fact, we will choose the numerical fluxes such 
that each individual interface term 
is non-positive. At interface $\mathsf{I}^{k+1/2}$ and
in $L,R$ notation, the jump and average of $\hat{v}_h$, for 
example, are
\begin{subequations}
\begin{align}
\frac{1}{2}\left(\hat{v}^+
+ \hat{v}^-\right)\equiv 
\lcurlyjump \hat{v}_h \rcurlyjump 
& = \frac{1}{2}\left(\hat{v}^L_{k+1/2} 
+ \hat{v}^R_{k+1/2}\right) 
\\
\mathbf{n}^- \hat{v}^-
+ \mathbf{n}^+\hat{v}^+
\equiv 
\ljump \hat{v}_h \rjump 
& =  \hat{v}^L_{k+1/2} - \hat{v}^R_{k+1/2}.
\end{align}
\end{subequations}
Consider numerical fluxes of the form
\begin{subequations} \label{eqn_nf_general}
\begin{align}
Q^* & = 
\lcurlyjump Q_h \rcurlyjump - \frac{\tau_{Q}}{2}\ljump Q_h \rjump  \\
\hat{v}{}^* & = 
\lcurlyjump \hat{v}_h \rcurlyjump 
- \frac{\tau_{v}}{2}\ljump \hat{v}_h \rjump  \\
\label{eqn_nf_u}
u^* & = 
\lcurlyjump u_h \rcurlyjump - \frac{\tau_{u}}{2}\ljump u_h \rjump \\
(\partial_t u)^* &  = 
\lcurlyjump \partial_t u_h \rcurlyjump 
- \frac{\tau_{u}}{2}\ljump \partial_t u_h \rjump,
\end{align}
\end{subequations}
where (\ref{eqn_nf_general}c) induces (\ref{eqn_nf_general}d) 
and where the penalty parameters $\tau_{u}$, $\tau_{v}$, and 
$\tau_{Q}$ are real numbers. 
The fluxes defined in \eqref{eqn:toyfluxes} 
correspond to $\tau_u = 1$, 
$\tau_v = 1+\sqrt{a}$, and $\tau_Q = 0$.
In terms of these quantities the $k$th interface contribution 
in \eqref{eq:ksummation} is
\begin{align}
\begin{split}
(\Delta^{L}_h & - \Delta^R_h) \big|_{\mathsf{I}^{k+1/2}}
= \frac{1}{2} \left( \ljump Q_h^2 \rjump 
- \ljump \hat{v}_h^2 \rjump \right) 
+ \lcurlyjump \hat{v}_h \rcurlyjump\ljump \hat{v}_h \rjump 
- \frac{\tau_v}{2}\ljump \hat{v}_h \rjump^2 \\
& +\sqrt{a}\lcurlyjump Q_h \rcurlyjump\ljump \hat{v}_h \rjump
- \frac{\sqrt{a}\tau_Q}{2}\ljump Q_h \rjump \ljump \hat{v}_h \rjump 
- \lcurlyjump Q_h \rcurlyjump\ljump \partial_t u_h \rjump
-\frac{\tau_u}{2}\ljump \partial_t u_h \rjump \ljump Q_h \rjump,
\end{split}
\end{align}
where we have suppressed the $k$ dependence of the right-hand side.
Now consider the term $\ljump \partial_t u_h \rjump$. Because 
$\partial_t u_h$ and $Q_h + \sqrt{a}\hat{v}_h$ are both 
polynomials of degree $N$, Eq.~(\ref{eqn:Toy_SD_WP}a) implies
the vector equation  $\partial_t\boldsymbol{u}
= \boldsymbol{Q} + \sqrt{a}\hat{\boldsymbol{v}}$, that is
{\em pointwise} equivalence on the nodal points of $\mathsf{D}^k$,
which in turn implies $\ljump \partial_t u_h\rjump = 
\ljump Q_h + \sqrt{a}\hat{v}_h\rjump$. Upon substituting 
this identity into the last equation, we arrive at an expression 
which features only $\hat{v}_h$ and $Q_h$,
\begin{align}
\begin{split} \label{eqn_interface_exp_2}
(\Delta^{L}_h & - \Delta^R_h) \big|_{\mathsf{I}^{k+1/2}}
= \frac{1}{2} \left( \ljump Q_h^2 \rjump 
- \ljump \hat{v}_h^2 \rjump \right) 
+ \lcurlyjump \hat{v}_h \rcurlyjump\ljump \hat{v}_h \rjump 
- \frac{\tau_v}{2}\ljump \hat{v}_h \rjump^2 \\
& +\sqrt{a}\lcurlyjump Q_h \rcurlyjump\ljump \hat{v}_h \rjump 
- \frac{\sqrt{a}\tau_Q}{2}\ljump Q_h \rjump \ljump \hat{v}_h \rjump 
- \lcurlyjump Q_h \rcurlyjump\ljump Q_h + \sqrt{a}\hat{v}_h \rjump
- \frac{\tau_u}{2}\ljump Q_h + \sqrt{a}\hat{v}_h \rjump \ljump Q_h \rjump.
\end{split}
\end{align}
The identities 
$\lcurlyjump \hat{v}_h \rcurlyjump\ljump \hat{v}_h \rjump 
= \frac{1}{2}\ljump \hat{v}_h^2 \rjump$ and 
$\ljump Q_h + \sqrt{a}\hat{v}_h \rjump=\ljump Q_h \rjump 
+ \sqrt{a}\ljump \hat{v}_h \rjump$ 
then simplify (\ref{eqn_interface_exp_2}) to
\begin{align}
(\Delta^{L}_h - \Delta^R_h) \big|_{\mathsf{I}^{k+1/2}} =
- \frac{\tau_v}{2}\ljump \hat{v}_h \rjump^2 
- \frac{\sqrt{a}(\tau_u +\tau_Q)}{2}
\ljump Q_h \rjump \ljump \hat{v}_h \rjump 
-\frac{\tau_u}{2}\ljump Q_h \rjump^2 .
\end{align}
The role of a penalty parameter is now clear. Positive
values of $\tau_v$ penalize jumps in $\hat{v}_h$ through 
a negative contribution to the energy. Likewise,
positive values of $\tau_u$ penalize jumps in $Q_h$ 
through a negative contribution to the energy. However,
because the sign of $\ljump Q_h \rjump \ljump \hat{v}_h \rjump$ 
can be positive or negative, only the choice
$\tau_Q = - \tau_u$ yields an expression
for $(\Delta^{L}_h - \Delta^R_h)|_{\mathsf{I}^{k+1/2}}$
which is manifestly negative for $\tau_u \geq 0$ 
and $\tau_v \geq 0$. A simple estimate 
based on Young's inequality with $\varepsilon$
(that is, $2\alpha\beta \leq \varepsilon^{-1}\alpha^2 
+ \varepsilon \beta^2$, where $\alpha,\beta \geq 0$ and 
$\varepsilon > 0$) shows that for $\tau_Q = 0$ the choice 
$\tau_v \geq a\tau_u/4$ also yields stability.
%%%%%%%%%%%%%%%%%%%%%%%%%%%%%%%%%%%%%%%%%%%%%%%%%%%%%%%%%%
% Figure for stability test.
\begin{figure}
\centering
\includegraphics[height=2.5in]{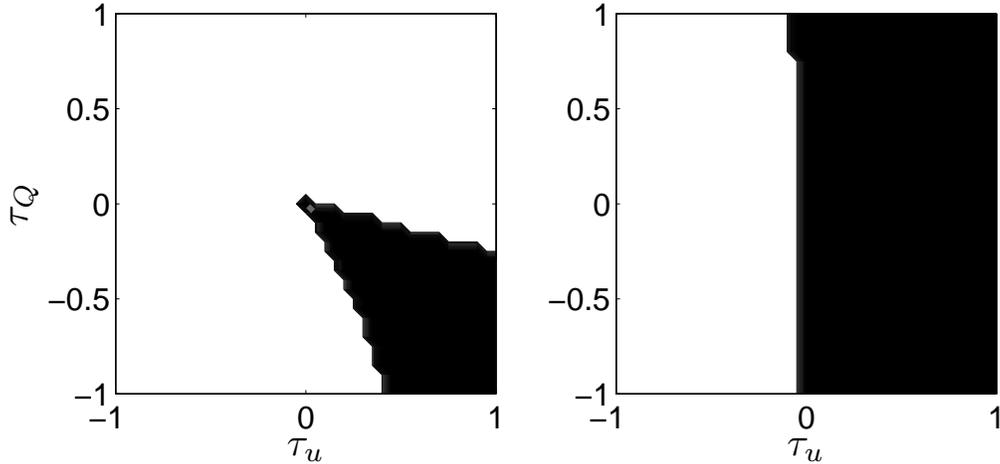}
\caption{{\sc Stable evolutions for the model system.}
For fixed $\tau_v = 10^{-6}$ and $\tau_v = 1 + \sqrt{2}$
respectively, the left and right plots depict stable choices
(determined empirically) of $\tau_u$ and $\tau_Q$ for the
linear model system \eqref{eqn:TOY_Lin}. The stable regions
are colored black, but the jagged edges result from the
discretization of the $(\tau_u,\tau_Q)$-plane.}
\label{fig:PenaltyStabilityTest}
\end{figure}

Figure \ref{fig:PenaltyStabilityTest} depicts 
certain choices of stable penalty parameters for
the {\em linear} model system evolved to
$t_{\mathrm{final}} = 1000$ (with $a=2$, $N = 10$, 
and $\Delta t \simeq 0.0553$), as determined 
empirically with simulations similar to those described 
in Sec.~\ref{subsec:toy_simulations}. The left plot
corresponds to a small $\tau_v = 10^{-6}$, for which
the choice $\tau_u = 1$, $\tau_Q = 0$ is not stable,
as expected from the theoretical analysis. However, 
the right plot corresponds to $\tau_v = 1 + \sqrt{a}$, 
for which $\tau_u = 1$, $\tau_Q = 0$ is stable. Motivated 
by the numerical flux choices
(\ref{eqn:f_fluxBSSN},\ref{eqn:alpha_fluxBSSN}) used for the
BSSN system \eqref{eqn:GBSSNL_spherical}, we have (as mentioned
above) set $\tau_u = 1$, $\tau_v = 1+\sqrt{a}$, and $\tau_Q = 0$ in
simulations of the nonlinear model \eqref{eqn:TOY}. 
For the nonlinear model system \eqref{eqn:TOY}, the theoretically
motivated choice $\tau_Q = - \tau_u$ also yields numerically
stable evolutions when $\tau_u \geq 0$ and $\tau_v \geq 0$.

For the nonlinear systems \eqref{eqn:GBSSNL_spherical}
and \eqref{eqn:TOY}, we do not attempt a formal stability proof.
Nevertheless, the results of this appendix have served as a 
guide for our choices of penalty parameters. For the BSSN system 
\eqref{eqn:GBSSNL_spherical}, $u$, $v$, and $Q$ are block indices 
[cf.~Eq.~\eqref{eqn:vectors_uvQ}]. Similar to the model problem, 
we have penalized $Q$ with $\tau_u = 1$, with $\tau_v$ 
chosen large enough to heuristically 
overcome the cross-terms of indefinite size that arise from 
$\tau_Q = 0$ (we interpret equations like $\tau_u = 1$ 
componentwise). An analogous choice 
``$\tau_Q = - \tau_u$" for the BSSN system might be possible, but 
would be considerably more complicated. Indeed, such a choice 
likely entails a {\em matrix} of penalty 
parameters, but we do not give the details here.

%%%%%%%%%%%%%%%%%%%%%%%%%%%%%%%%%%%%%%%%%%%%%%%%%%%%%%%%%%%%%%%
%
%  REFERENCES CITED
%
%%%%%%%%%%%%%%%%%%%%%%%%%%%%%%%%%%%%%%%%%%%%%%%%%%%%%%%%%%%%%%%

\bibliographystyle{plain}
\end{document}